\documentclass[a4paper, 12pt]{article}

\usepackage{styleBuding}
\usepackage{bm}
\usepackage{color,listings}
\usepackage[usenames,dvipsnames,svgnames,table]{xcolor}
\usepackage{amsthm, amsmath}
\usepackage{amssymb}
\usepackage{mathrsfs}
\usepackage{graphicx}
\usepackage{slashed}
\usepackage{soul}
\usepackage{multirow}
\usepackage{subfigure}
\usepackage{diagbox}
\usepackage{siunitx} 
\usepackage{url} 
\usepackage[utf8]{inputenc}
\usepackage[figuresright]{rotating}
\definecolor{back}{HTML}{F8F8F8}
\usepackage{epstopdf}
\usepackage{lipsum}

\usepackage{booktabs}
\usepackage{array}

\newcommand{\rom}[1]{\uppercase\expandafter{\romannumeral #1\relax}}

\allowdisplaybreaks

\title{Constraining the light Higgs bosons in the GNMSSM with recent Higgs data}
\author{Zhaoxia Heng, Zehan Li, Haijing Zhou}
\affiliation{ School of Physics, Henan Normal University, Xinxiang 453007, China}
\emailAdd{zxheng@htu.edu.cn}
\emailAdd{hnuzehanli@163.com}
\emailAdd{zhouhaijing0622@163.com}

\abstract{
The search for light scalar and pseudoscalar particles  provides a promising avenue for probing physics beyond the Standard Model (SM).
In this study, we investigated the exotic decay channels of the 125 GeV SM-like Higgs boson into pairs of light CP-odd ($a_s$) or CP-even ($h_s$) Higgs bosons within the framework of the General Next-to-Minimal Supersymmetric Standard Model (GNMSSM). 
A comprehensive parameter space scan is performed using the MultiNest algorithm, incorporating constraints from $ \textsf{HiggsSignals-2.6.2} $, $ \textsf{HiggsBounds-5.10.2} $, and ATLAS experimental searches, under two distinct scenarios 
where either the lightest ($h_1$) or next-to-lightest ($h_2$) CP-even state is the observed Higgs boson ($h$). 
Our results demonstrate that $ \textsf{HiggsBounds} $ imposes the most stringent exclusion limits due to its sensitivity to direct searches for non-SM Higgs bosons.
In the $h_2$ scenario, $ \textsf{HiggsSignals} $ can additionally exclude regions with suppressed exotic branching ratios (e.g., $ Br(h \to a_sa_s \to \tau\tau bb) \leq 2.5\%$), due to its sensitivity to indirect deviations caused by the kinematically enhanced decay \( h \to h_sh_s \). 
Under combined constraints from \( \textsf{HiggsTools} \), $h$ must retain at least 93\% SM-like component (\( V_h^\text{SM} \geq 0.93 \)) with no more than 32\% singlet admixture (\( V_h^\text{S} \leq 0.32 \)); in the $ h_2 $ case, the lightest scalar $ h_s $ exhibits high singlet purity ($ V_{h_s}^\text{S} \geq 0.94 $). Furthermore, dark matter (DM) phenomenology indicates that singlino- or higgsino-dominated DM is viable in the $ h_1 $ scenario, with dominant annihilation channels including $ \tilde{\chi}_1^0\tilde{\chi}_1^0 \to h_sa_s $ for singlino-like DM and chargino co-annihilation for higgsino-like DM, whereas the $ h_2 $ scenario favors higgsino-dominated DM.
}

\begin{document}
    \maketitle
    \flushbottom

\section{Introduction}

The discovery of the 125 GeV Higgs boson at the Large Hadron Collider (LHC) \cite{ATLAS:2012ae,Chatrchyan:2012tx} marked a milestone in particle physics because it confirmed the electroweak symmetry breaking (EWSB) mechanism described by the Standard Model (SM). However, the SM leaves several fundamental problems unsolved, such as the neutrino masses, hierarchy problem, and dark matter (DM) problem. These unresolved issues have motivated exploration of physics beyond the Standard Model (BSM), where extended Higgs sectors and new symmetries often predict additional scalar states.

Supersymmetry (SUSY) is a  well-motivated candidate for BSM, providing solutions to the hierarchy problem and gauge coupling unification. The Minimum Supersymmetric Standard Model (MSSM) \cite{Nilles:1983ge,Haber:1984rc,Gunion:1984yn,Martin:1997ns}, as the most economical SUSY realization, has been extensively studied. However, it is challenged by DM constraints, the $\mu$ problem, the little hierarchy problem, and other issues. As a natural and minimal extension, the Next-to-Minimal Supersymmetric Standard Model (NMSSM) \cite{Miller:2003ay,Ellwanger:2009dp,Maniatis:2009re} introduces a gauge-singlet Higgs superfield. Its scalar component acquires a non-zero vacuum expectation value (vev), thereby generating an effective $\mu$-term. Mixing between singlet and doublet superfields produces an additional CP-even Higgs ($h_s$) that mixes with MSSM-like states. This mixing can suppress the lightest Higgs mass while enabling the next-to-lightest Higgs to match the observed 125 GeV signal \cite{Cao:2012fz,Cao:2018rix,Cao:2016nix,Cao:2016cnv,Heng:2018kyd,Cao:2016uwt,Cao:2014kya}. 
This scenario alleviates fine-tuning and offers rich phenomenology, including Higgs-to-Higgs decays \cite{Ellwanger:2024etv,Ellwanger:2022jtd,Ma:2020mjz,Wang:2016lvj,Cao:2013gba,Curtin:2013fra,Cheng:2024gfs,Zhou:2025xol,Liu:2016ahc,Domingo:2016unq,Curtin:2014pda,King:2014xwa,Liu:2013gea} or exotic decays that involve neutralinos \cite{Huang:2013ima,Huang:2014cla}.

Over the past decade, the NMSSM with the $Z_3$-symmetric ($Z_3$-NMSSM) 
\cite{Miller:2003ay,Ellwanger:2009dp,Maniatis:2009re} has drawn significant interest as a simple extension of the MSSM. 
However, experimental searches for light Higgs bosons, electroweakinos, and extra scalars have excluded a substantial portion of the theoretically natural parameter space, especially  in regions where the 125 GeV Higgs mass, B-physics, and DM relic density are simultaneously satisfied \cite{Cao:2022htd,Zhou:2021pit}. 
In addition, LHC searches for supersymmetric particles and tighter limits from DM direct detection experiments such as XENON-nT~\cite{XENON:2018voc}, PandaX-4T~\cite{PandaX-II:2020oim,PandaX-II:2017hlx}, and LUX-ZEPLIN (LZ)~\cite{LZ:2022lsv,LZ:2024zvo}, have severely constrained the viable parameter space of the NMSSM.
For large $\lambda$, predicted spin-independent scattering cross sections approach current experimental sensitivity~\cite{Cao:2022htd,Zhou:2021pit}; only  scenarios with small $\lambda$ or destructive cancellations can evade these bounds. However, small $\lambda$ makes it difficult to achieve the correct Higgs mass without significant fine-tuning~\cite{Cao:2022htd,Zhou:2021pit}.
Consequently,  the $Z_3$-NMSSM survives only in a narrow and fine-tuned region, challenging its original goal of solving the naturalness problem.

To address the aforementioned theoretical limitations of the $Z_3$-NMSSM concerning  the Higgs boson mass, the superparticle mass spectrum, and its connection to 
DM, researchers have proposed various extended frameworks, among which the General NMSSM (GNMSSM) \cite{Choi:2019yrv,Cao:2022ovk,Cao:2024axg,Meng:2024lmi,Cao:2021ljw,95GeV2023,Cao:2022chy} stands out as a representative example. 
By adding extra singlet couplings or $Z_3$-symmetry-breaking terms, the GNMSSM enhances  flexibility in the Higgs mass matrix and enables lighter CP-even and CP-odd states, which are consistent with LHC data \cite{Yue:2025dqe}.
In the framework of the GNMSSM, the lightest supersymmetric particle (LSP)  serves as a viable DM candidate. 
Singlet-dominant particles such as singlino-dominated DM and singlet-dominated Higgs bosons, can form a secluded DM sector \cite{Pospelov:2007mp, Meng:2024lmi}. In this sector, singlino-like DM primarily annihilates via the s-channel exchange of light CP-even or CP-odd Higgs bosons. 
The coupling strength  between these scalar  particles and DM  governs  both the DM relic abundance from the early universe and the scattering cross-section with nucleons in  direct detection experiments. This connection establishes a link between collider physics and DM phenomenology, enabling constraints from high-energy and astrophysical observations to complement each other.

The couplings between these light Higgs bosons and SM fermions or gauge bosons are dynamically suppressed, leading to very low production rates in diphoton, ZZ, or WW channels. Consequently,  these particles are difficult to detect with conventional search strategies. However, they can still appear indirectly through hidden-sector effects, especially via exotic Higgs decays.
In parameter  regions where the light Higgs bosons are lighter than half of the 125 GeV Higgs boson $h$, the latter can decay into pairs of such Higgs (e.g., $h\to h_sh_s$ or $a_sa_s$ ), each of which subsequently decays into b-jets, $\tau$-lepton pairs, or other SM states. Although these cascade decays have small branching ratios, their low backgrounds make them sensitive probes for new physics. Meanwhile, the Higgs coupling to a light singlet scalar particle ($m_s < m_h/2$)  drives  a strong  first-order electroweak phase transition (SFOEWPT) ~\cite{Profumo:2007wc,Kozaczuk:2019pet,Carena:2019une,Carena:2022yvx}. Therefore, these light Higgs bosons are a compelling target for exotic Higgs decay searches at the LHC and future colliders. ATLAS and CMS have systematically searched for such signatures across multiple LHC runs, targeting final states such as $4b$ \cite{CMS:2024zfv}, $\mu\mu bb$ \cite{CMS2024,CMS:2017dmg,ATLAS:2018emt}, $\tau\tau bb$ \cite{CMS:2018zvv,ATLAS2024}, $\gamma\gamma \tau\tau$ \cite{ATLAS:2024nnm}, and $4\mu$ \cite{ATLAS:2015hpr,CMS:2012qms,CMS:2015nay,ATLAS:2018coo,CMS:2018jid}.

This study investigates the $\tau\tau bb$ final state~\cite{ATLAS2024}. Compared with the $4b$ channel \cite{CMS:2024zfv}, this signature improves background suppression by exploiting the distinctive decay topology of the $\tau$ lepton, while retaining high signal efficiency. The \textsf{HiggsTools} package~\cite{Bahl:2022igd} is employed to constrain the model parameter space, identifying regions consistent with direct search limits and Higgs property measurements. Within the GNMSSM framework, we then examine the exotic decays $h\to h_sh_s / a_sa_s$ and evaluate their detectability. A preliminary analysis of DM properties in the allowed parameter region is also presented. 

This paper is organized as follows. Section 2 introduces the GNMSSM framework and describes in detail its Higgs sector and the key trilinear Higgs couplings. Section 3 outlines our scanning strategy and the corresponding numerical results. Section 4 summarizes our conclusions.

 \section{\label{theory-section}Theoretical preliminaries}
\subsection{\label{Section-Model}Basics of the GNMSSM }
The GNMSSM extends the MSSM by introducing a gauge-singlet superfield $\hat{S}$ that carries no baryon or lepton number. Consequently, the Higgs sector consists of the standard $ SU(2)_L$ doublets \( \hat{H}_u = (\hat{H}^+_u, \hat{H}^0_u) \) and \( \hat{H}_d = (\hat{H}^0_d, \hat{H}^-_d) \)  , together with the singlet $\hat{S}$.
Its gauge-invariant superpotential can be written as ~\cite{Ellwanger:2009dp}:
\begin{eqnarray}
W_{\rm GNMSSM} = W_{\text{Yukawa} }+ \lambda \hat{S}\hat{H_u} \cdot \hat{H_d} + \frac{\kappa}{3}\hat{S}^3 + \mu \hat{H_u} \cdot \hat{H_d} + \frac{1}{2} \mu^{\prime} \hat{S}^2 + \xi\hat{S}, \label{Superpotential}
\end{eqnarray}
where $W_{\rm Yukawa}$ contains the MSSM quark and lepton Yukawa interactions.
The dimensionless couplings $\lambda$ and $\kappa$ govern the Higgs interactions, which are analogous to the $Z_3$-NMSSM.  The GNMSSM incorporates explicit $Z_3$-symmetry-breaking via parameters $\mu$, $\mu^\prime$ and $\xi$.
These terms resolve the inherent tadpole problem \cite{Ellwanger:1983mg, Ellwanger:2009dp} and cosmological domain-wall problem \cite{Abel:1996cr, Kolda:1998rm, Panagiotakopoulos:1998yw} in the $Z_3$-NMSSM.
Through a redefinition of the singlet field \cite{Ross:2011xv}, the parameter $\xi$ can be absorbed, allowing it to be set to zero without loss of generality. 
$\mu$ and $\mu^\prime$ arise from the spontaneous breaking of discrete $R$-symmetries ($\mathbb{Z}^R_4$ or $\mathbb{Z}^R_8$) at high scales \cite{Abel:1996cr, Lee:2010gv, Lee:2011dya, Ross:2011xv, Ross:2012nr}.
The inclusion of these explicit $Z_3$-breaking terms modifies the neutral Higgs mass spectrum and generates markedly richer phenomenology than the $Z_3$-NMSSM and MSSM.

\subsection{\label{DMRD}Higgs sector of the GNMSSM}

The soft SUSY-breaking terms in the Higgs sector of the GNMSSM can be expressed as:
\begin{eqnarray}
    -\mathcal{L}_{soft} = &\Bigg[\lambda A_{\lambda} S H_u \cdot H_d + \frac{1}{3} \kappa A_{\kappa} S^3+ m_3^2 H_u\cdot H_d + \frac{1}{2} {m_S^{\prime}}^2 S^2 + \xi^\prime S + h.c.\Bigg] \nonumber \\
   & + m^2_{H_u}|H_u|^2 + m^2_{H_d}|H_d|^2 + m^2_{S}|S|^2, \label{Soft-terms}
     \end{eqnarray}
where \( H_u, H_d, \) and \( S \) are scalar components of the Higgs superfields,
and \( m^2_{H_u}, m^2_{H_d}, \) and \( m^2_S \) are their supersymmetry-breaking masses. After the electroweak symmetry breaking, the neutral components of the Higgs fields develop non-vanishing vevs:
\begin{eqnarray}
\langle H^0_u \rangle = \frac{v_u}{\sqrt{2}}, \quad \langle H^0_d \rangle = \frac{v_d}{\sqrt{2}}, \quad \langle S \rangle = \frac{v_s}{\sqrt{2}}, \label{2.3}
     \end{eqnarray}
with \( v = \sqrt{v_u^2 + v_d^2} \simeq 246 \, \text{GeV} \). Then, the Higgs sector is characterized by 11 independent physical parameters:
\begin{eqnarray}
\tan \beta \equiv \frac{v_u}{v_d}, \, \lambda, \, \kappa, \, v_s, \, A_\lambda, \, A_\kappa, \, \mu, \, \mu', \, m_3^2, \, m_S^{\prime 2}, \, \xi'. \label{2.4}   \end{eqnarray}

To investigate the phenomenology of the Higgs sector, it is convenient to employ the following specialized parametrization:
\begin{eqnarray}
H_{\rm NSM} \equiv \cos\beta {\rm Re}(H_u^0) - \sin\beta {\rm Re}(H_d^0), \nonumber\\
H_{\rm SM} \equiv \sin\beta {\rm Re}(H_u^0) + \cos\beta {\rm Re} (H_d^0),\\
A_{\rm NSM} \equiv \cos\beta {\rm Im}(H_u^0) - \sin\beta {\rm Im}(H_d^0)\nonumber.
\end{eqnarray}
 In the basis of ($H_{\rm NSM}$, $H_{\rm SM}$, ${\rm Re[S]}$), the mass matrix of CP-even Higgs fields can be written as:~\cite{Meng:2024lmi}

\begin{eqnarray}
\mathcal{M}^2_{S,11} &=& \frac{\lambda v_s(\sqrt{2}A_\lambda + \kappa v_s + \sqrt{2}\mu') + 2m^2_3}{\sin 2\beta} + \frac{1}{2}(2m^2_Z - \lambda^2 v^2) \sin^2 2\beta, \nonumber\\
\mathcal{M}^2_{S,12} &=& -\frac{1}{4}(2m^2_Z - \lambda^2 v^2) \sin 4\beta, \quad \mathcal{M}^2_{S,13} = -\frac{\lambda v}{\sqrt{2}} \left( A_\lambda + \sqrt{2} \kappa v_s + \mu' \right) \cos 2\beta,\nonumber\\
\mathcal{M}^2_{S,22} &=& m^2_Z \cos^2 2\beta + \frac{1}{2} \lambda^2 v^2 \sin^2 2\beta, \nonumber\\
\mathcal{M}^2_{S,23} &=& \frac{\lambda v}{\sqrt{2}} \left[ \left( \sqrt{2}\lambda v_s + 2\mu \right) - (A_\lambda + \sqrt{2}\kappa v_s + \mu') \sin 2\beta \right], \nonumber\\
\mathcal{M}^2_{S,33} &=&  \frac{(A_\lambda + \mu')\sin2\beta}{2\sqrt{2}v_s} \lambda v^2 + \frac{\kappa v_s}{\sqrt{2}}(A_\kappa + 2\sqrt{2}\kappa v_s + 3\mu')  - \frac{\mu}{\sqrt{2}v_s} \lambda v^2 - \frac{\sqrt{2}}{v_s} \xi'.
 \label{2.5}
     \end{eqnarray}

Similarly, based on ($A_{\rm NSM}$, ${\rm Im[S]}$), the mass matrix of CP-odd Higgs fields can be written as \cite{Meng:2024lmi}
\begin{eqnarray}
\mathcal{M}^2_{P,11} &=& \frac{\lambda v_s \left( \sqrt{2} A_\lambda + \kappa v_s + \sqrt{2} \mu' \right) + 2 m^2_3}{\sin 2\beta},\quad \mathcal{M}^2_{P,12} = \frac{\lambda v}{\sqrt{2}} \left( A_\lambda - \sqrt{2} \kappa v_s - \mu' \right), \nonumber\\
\mathcal{M}^2_{P,22} &=& \frac{(A_\lambda + 2 \sqrt{2} \kappa v_s + \mu') \sin 2\beta}{2 \sqrt{2} v_s} \lambda v^2 - \frac{\kappa v_s}{\sqrt{2}}(3 A_\kappa + \mu') \nonumber\\
&-&\frac{\mu}{\sqrt{2} v_s} \lambda v^2 - 2 m'^2_S - \frac{\sqrt{2}}{v_s} \xi'.\label{CP-odd-mass-matrix}
\end{eqnarray}

The three CP-even mass eigenstates $h_i$ ($i=1,2,3$) and the two CP-odd Higgs mass eigenstates $a_j$ ($j=1,2$) are obtained by diagonalizing the scalar ($\mathcal{M}^2_{S}$) and pseudoscalar ($\mathcal{M}^2_{P}$) mass matrices, respectively:
\begin{eqnarray} \label{Mass-eigenstates}
  h_i & = & V_{h_i}^{\rm NSM} H_{\rm NSM}+V_{h_i}^{\rm SM} H_{\rm SM}+V_{h_i}^{\rm S} Re[S], \nonumber \\
  a_j & = &  V_{P, a_j}^{\rm NSM} A_{\rm NSM}+ V_{P, a_j}^{\rm S} Im [S].
\end{eqnarray}
For convenience, the mass of three CP-even states and two CP-odd states are ordered by increasing mass: $m_{h_1} < m_{h_2} < m_{h_3}$ and $m_{a_1} < m_{a_2}$. $|V_{h_i}^{\rm NSM}|^2$, $|V_{h_i}^{\rm SM}|^2$, and $|V_{h_i}^{\rm S}|^2$ ($|V_{h_i}^{\rm NSM}|^2$+$|V_{h_i}^{\rm SM}|^2$+ $|V_{h_i}^{\rm S}|^2$=1) denote the composition ratios of the non-SM scalar component $H_{\rm NSM}$, the SM component $H_{\rm SM}$, and the singlet scalar component $\mathrm{Re}[S]$ in the $h_i$, respectively. Similarly, $|V_{P, a_j}^{\rm NSM}|^2$ and $|V_{P, a_j}^{\rm S}|^2$ denote the composition ratios of the non-SM pseudoscalar component $A_{\rm NSM}$ and the singlet pseudoscalar component $\mathrm{Im}[S]$ in the CP-odd mass eigenstate $a_j$, respectively. 
In terms of dominant components, the physical states are labeled as follows:
\begin{itemize}
\item $h$: the physical Higgs state with $|V_{h}^{\rm SM}|^2 > 0.5$, corresponding to the SM-like Higgs boson.
\item $H$, $A_H$: the CP-even and CP-odd non-SM-like Higgs states with $|V_{H}^{\rm NSM}|^2 > 0.5$ and $|V_{P, A_H}^{\rm NSM}|^2 > 0.5$, respectively.
\item $h_s$, $a_s$: the scalar and pseudoscalar singlet-like Higgs states with  $|V_{h_s}^{\rm S}|^2 > 0.5$ and $|V_{P, a_s}^{\rm S}|^2 > 0.5$, respectively.
\end{itemize}
The identity of the 125 GeV Higgs depends on the mass hierarchy:
\begin{itemize}
    \item $h_1$ scenario: $h \equiv h_1$ with singlet-dominated $h_s$ heavier ($m_{h_s} > m_h$)
    \item $h_2$ scenario: $h \equiv h_2$ with singlet-dominated $h_s$ lighter ($m_{h} > m_{h_s}$)
\end{itemize}
In addition, the model predicts a pair of charged Higgs bosons:
\begin{eqnarray}
 H^\pm = \cos \beta H_u^\pm + \sin \beta H_d^\pm,
\end{eqnarray}
 with the following mass: \cite{Ellwanger:2009dp,Meng:2024lmi}
\begin{eqnarray}
    m^2_{H^\pm} = \frac{\lambda v_s \left( \sqrt{2} A_\lambda + \kappa v_s + \sqrt{2} \mu' \right) + 2 m^2_3}{\sin 2\beta} + m^2_W - \frac{1}{2} \lambda^2 v^2. \label{Charged Hisggs Mass}
  \end{eqnarray}

To connect the theoretical inputs with experimental observables, we replaced the Lagrangian parameters
 \( \mu, \mu', m^2_3, m'^2_S, \) and \( \xi' \) with physical mass parameters:

\begin{itemize}
    \item $m_A\equiv \sqrt{M^2_{P,11}}$: mass of a heavy MSSM-like CP-odd Higgs boson,
\item $m_B \equiv \sqrt{M^2_{S,33}}$: mass of a CP-even singlet Higgs boson,
\item $m_C \equiv \sqrt{M^2_{P,22}}$: mass of a CP-odd singlet Higgs boson,
\item $\mu_{\text{tot}} \equiv \mu_{\text{eff}} + \mu$: higgsino mass parameter,
\item $m_N \equiv \frac{2\kappa}{\lambda} \mu_{\text{eff}} + \mu'$: singlino mass parameter.
\end{itemize}
Then, the original Lagrangian parameters become:
\begin{eqnarray}
\mu &=& \mu_{\text{tot}} - \frac{\lambda}{\sqrt{2}} v_s, \quad \mu' = m_N - \sqrt{2} \kappa v_s, \quad m^2_3 = \frac{m^2_A \sin 2\beta}{2} - \lambda v_s \left( \frac{\kappa v_s}{2} + \frac{\mu'}{\sqrt{2}} + \frac{A_\lambda}{\sqrt{2}} \right), \nonumber\\
\xi' &=& \frac{v_s}{\sqrt{2}} \left[ \frac{(A_\lambda + \mu') \sin 2\beta}{2\sqrt{2} v_s}\lambda v^2 + \frac{\kappa v_s}{\sqrt{2}} (A_\kappa + 2\sqrt{2} \kappa v_s + 3\mu')  - \frac{\mu}{\sqrt{2} v_s} \lambda v^2 - m^2_B \right], \nonumber\\
m'^2_S &=& \frac{1}{2} \left[ m^2_B - m^2_C + \lambda \kappa \sin 2\beta v^2 - 2\sqrt{2} \kappa v_s (A_\kappa + \frac{\kappa}{\sqrt{2}} v_s + \mu') \right]. \label{2.9}
  \end{eqnarray}
This reparameterization simplifies the Higgs mass matrices (Eqs. \ref{2.5}, \ref{CP-odd-mass-matrix}) to:
\begin{eqnarray}
\mathcal{M}^2_{S,11} &=& m^2_A + \frac{1}{2}(2m^2_Z - \lambda^2 v^2) \sin^2 2\beta, \quad \mathcal{M}^2_{S,12} = -\frac{1}{4}(2m^2_Z - \lambda^2 v^2) \sin 4\beta, \nonumber\\
\mathcal{M}^2_{S,13} &=& -\frac{\lambda v}{\sqrt{2}}(A_\lambda + m_N) \cos 2\beta, \quad \mathcal{M}^2_{S,22} = m^2_Z \cos^2 2\beta + \frac{1}{2} \lambda^2 v^2 \sin^2 2\beta, \nonumber\\
\mathcal{M}^2_{S,23} &=& \frac{\lambda v}{\sqrt{2}}[2\mu_{\text{tot}} - (A_\lambda + m_N) \sin 2\beta], \quad \mathcal{M}^2_{S,33} = m^2_B, \nonumber\\
\quad \mathcal{M}^2_{P,11} &=& m^2_A, \quad
\mathcal{M}^2_{P,22} = m^2_C, \quad \mathcal{M}^2_{P,12} = \frac{\lambda v}{\sqrt{2}}(A_\lambda - m_N).\label{2.10}
  \end{eqnarray}
The neutral Higgs sector is fully specified by eight parameters:
$\tan \beta$, $\lambda$, $A_\lambda$, $m_A$, $m_B$, $m_C$, $m_N$, and $\mu_{tot}$.
 The remaining parameters ($\kappa$, $A_\kappa$, and $v_s$) exclusively govern triple Higgs couplings \cite{Meng:2024lmi}.
 Considering the dominant radiative corrections from the top/stop loops to the mass of the SM-like Higgs boson, the mass of $h$ can be approximated as \cite{Ellwanger:2009dp, Cao:2012fz}
\begin{eqnarray}
m_h^2 & \simeq & m^2_Z \cos^2 2\beta + \frac{1}{2} \lambda^2 v^2 \sin^2 2\beta + \Delta^2
  \label{h-mass}   \end{eqnarray} 
with
\begin{eqnarray}
\Delta^2 &\simeq & \frac{3m_t^4}{2\pi^2 v^2} \left[\ln\frac{M_S^2}{m_t^2}+ \frac{X_t^2}{M_S^2}\left(1-\frac{X_t^2}{12M_S^2}\right)\right ]
     \end{eqnarray}  
where $M_S=\sqrt{m_{\tilde{t}_1 \tilde{t}_2}}$, $m_{\tilde{t}_1}$ and $m_{\tilde{t}_2}$ are the masses of the two stop eigenstates, and $X_t = A_t - \mu_{\text{tot}}\cot\beta $.

In the light $h_s$ scenario, the scalar mass matrix element ${\cal M}^2_{S, 23}$ is parameterized as:
\begin{eqnarray}
    {\cal M}^2_{S, 23} = \sqrt{2} \lambda v \delta \mu_{\rm tot}
    \label{M23}
\end{eqnarray}
where the dimensionless parameter $\delta \equiv [ 2 \mu_{\rm tot} - (A_\lambda + m_N) \sin 2 \beta]/(2 \mu_{\rm tot})$ quantifies the cancellation between effective $\mu$-term and SUSY-breaking contributions. This parametrization offers advantages: it naturally accommodates larger $\lambda$ couplings while remaining compatible with LHC Higgs data through modest $\delta$ values. Thus, we used $\delta$ as an input parameter instead of $A_\lambda$ in this work.

In the scenario with very heavy charged Higgs bosons, approximate expressions can be derived for the singlet-dominated states and their mixing parameters~\cite{Baum:2017enm}:
\begin{eqnarray}
m_{h_s}^2 & \simeq & m_B^2 - \frac{{\cal M}^4_{S, 13}}{m_A^2 - m_B^2}, \quad m_{a_s}^2 \simeq m_C^2 - \frac{{\cal M}^4_{P, 12}}{m_A^2 - m_C^2}, \quad \frac{V_{P,a_s}^{\rm NSM}}{V_{P,a_s}^{\rm S}} = \frac{{\cal M}^2_{P, 12}}{m_{a_s}^2 - m_A^2} \simeq 0, \nonumber \\
\frac{V_{h}^{\rm S}}{V_h^{\rm SM}} & \simeq & \frac{{\cal M}^2_{S, 23}}{m_h^2 - m_B^2}, \quad V_{h}^{\rm NSM} \sim 0, \quad V_h^{\rm SM} \simeq \sqrt{1 - \left ( \frac{V_{h}^{\rm S}}{V_h^{\rm SM}} \right )^2} \sim 1, \nonumber  \\
\frac{V_{h_s}^{\rm SM}}{V_{h_s}^{\rm S}} & \simeq &  \frac{{\cal M}^2_{S, 23}}{m_{h_s}^2 - m_h^2}, \quad V_{h_s}^{\rm NSM} \sim 0, \quad V_{h_s}^{\rm S} \simeq \sqrt{1 - \left ( \frac{V_{h_s}^{\rm SM}}{V_{h_s}^{\rm S}} \right )^2 } \sim 1.   \label{Approximations}
\end{eqnarray}
These expressions indicate that $V_h^S \simeq - V_{h_s}^{\rm SM}$ and that all scale with $\lambda$.
In the limit of $\lambda \to 0$, the singlet fields decouple from the doublet Higgs fields, and the masses $m_B$, $m_C$, and $m_N$ can be treated as the physical particle
masses with high accuracy. Moreover, the singlet masses are independent
and may assume small values, as they are only weakly constrained by current experimental data.

\subsection{Neutralino sector of the GNMSSM}

The interaction between gauginos and the fermionic components of the neutral Higgs bosons generates five neutralinos and two charginos, represented as $\tilde{\chi}_i^0$ ($i=1,\dots,5$) and $\tilde{\chi}_i^\pm$ ($i=1,2$), respectively.
In the gauge eigenstate basis $\psi^0 = \left(-i \tilde{B}, -i\tilde{W}, \tilde{H}_d^0, \tilde{H}_u^0, \tilde{S}\right)$, the corresponding symmetric neutralino mass matrix is given by~\cite{Ellwanger:2009dp}
\begin{equation}
    M_{\tilde{\chi}^0} = \left(
    \begin{array}{ccccc}
    M_1 & 0 & -m_Z \sin \theta_W \cos \beta & m_Z \sin \theta_W \sin \beta & 0 \\
      & M_2 & m_Z \cos \theta_W \cos \beta & - m_Z \cos \theta_W \sin \beta &0 \\
    & & 0 & -\mu_{\text{tot}} & - \frac{1}{\sqrt{2}} \lambda v \sin \beta \\
    & & & 0 & -\frac{1}{\sqrt{2}} \lambda v \cos \beta \\
    & & & & m_{\text{N}}
    \end{array}
    \right), \label{eq:mmn}
\end{equation}
where $M_1$ and $M_2$ denote the gaugino soft-breaking mass parameters, while $s_W \equiv \sin \theta_W$ and $c_W \equiv \cos \theta_W$.
By diagonalization the mass matrix using the rotation matrix $N$, the physical neutralino mass eigenstates $\tilde{\chi}^0_i$ are obtained as follows:
\begin{eqnarray} \label{Mass-eigenstate-neutralino}
    \tilde{\chi}_i^0 = N_{i1} \psi^0_1 + N_{i2} \psi^0_2 + N_{i3} \psi^0_3 + N_{i4} \psi^0_4 + N_{i5} \psi^0_5,
\end{eqnarray}
where the index $i=1,2,...,5$, arranged in ascending order of mass. The lightest neutralino, $\tilde{\chi}_1^0$, can serve as the DM candidate. In this state, $N_{13}^2 + N_{14}^2$ and $N_{15}^2$ correspond to the higgsino and singlino components of the physical state $\tilde{\chi}_1^0$, respectively.
The lightest neutralino $\tilde{\chi}_1^0$ is classified as higgsino-dominated DM if $( N_{13}^2 + N_{14}^2 ) > 0.5$
and singlino-dominated DM if $N_{15}^2 > 0.5$.
In scenarios where gauginos are significantly heavier and the condition $\mu_{tot}^2-m_N^2 >> \lambda^2 v^2$ holds, the mass of the singlino-dominated DM can be approximately given by \cite{Meng:2024lmi}
\begin{eqnarray} \label{lightest neutralino mass}
m_{\tilde{\chi}_1^0} \simeq m_N + \frac{\lambda^2 v^2 (m_{\tilde{\chi}_1^0}- \mu_{tot}\sin 2\beta)}{2(m_{\tilde{\chi}_1^0}^2- \mu_{tot}^2)}\simeq m_N
\end{eqnarray}

\subsection{\label{coupling}Trilinear Higgs couplings in the GNMSSM}

For the light Higgs scenario in this study, we assumed a light CP-odd Higgs boson ($a_1$) that satisfies \( 2m_{a_1} < m_h \) to enable the exotic decay $h\to a_1a_1$ (i.e. $h\to a_sa_s$) in both $h_1$ and $h_2$ scenarios.
For the $h_2$ scenario, where the lightest CP-even Higgs boson $h_1$ is singlet-dominant, $h\to h_1h_1$ (i.e., $h\to h_sh_s$) becomes accessible when $2m_{h_1} < m_h$.
Both ATLAS and CMS collaborations have conducted searches  for non-SM Higgs bosons (\( H \), \( h_s \), \( A_H \) and \( a_s \)), which established stringent exclusion bounds on key parameters such as the masses and relative couplings \cite{ATLAS:2020zms,ATLAS:2021upq}.

The branching ratios of the exotic decays $h\to a_1a_1$ and $h\to h_1h_1$ are governed by their respective trilinear Higgs couplings $C_{ha_1a_1}$($\equiv C_{ha_sa_s}$) and $C_{hh_1h_1}$ ($\equiv C_{hh_sh_s}$). In the GNMSSM, these couplings are relevant to vacuum expectation values ($v_u$, $v_d$, $v_s$) and soft SUSY-breaking trilinear terms ($A_\lambda$, $A_\kappa$). While general analytic expressions for these couplings are provided in Ref.~\cite{Ellwanger:2022jtd}, the specific trilinear couplings relevant to our discussions are explicitly defined  as follows \cite{Meng:2024lmi}:
\begin{eqnarray}
 C_{h h_s h_s} &=& \lambda v V_{h}^{\rm SM} V_{h_s}^S V_{h_s}^{\rm S}
 (\lambda-\kappa \sin 2\beta) -\lambda \kappa v V_{h}^{\rm NSM} V_{h_s}^{\rm S} V_{h_s}^{\rm S} \cos 2\beta \nonumber \\
&& + \sqrt{2}\kappa V_{h}^{\rm S} V_{h_s}^{\rm S} V_{h_s}^{\rm S} (3m_N+A_\kappa)+
C_{h_s}^{\prime}(\lambda, \kappa, \tan\beta, v_s, A_{\lambda}, m_N), \label{haa}\\
 C_{h a_s a_s} &=& \lambda v V_{h}^{\rm SM} (\lambda+\kappa \sin 2\beta)+\lambda \kappa v V_{h}^{\rm NSM} \cos2\beta \nonumber \\
&& + \sqrt{2}\kappa V_{h}^{\rm S} (m_N-A_\kappa)+
C_{a_s}^{\prime}(\lambda, \kappa, \tan\beta, v_s, A_{\lambda}, m_N),
\label{hhh}
\end{eqnarray}
where the last terms in the two equations, suppressed by Higgs mixings, depend on $\lambda, \kappa, \tan\beta, v_s, A_{\lambda}$ and $m_N$.

\section{\label{numerical study}Numerical result}

\begin{table}[tbp]
  \centering
  \vspace{0.3cm}
  \resizebox{0.7\textwidth}{!}{
  \begin{tabular}{c|c|c|c|c|c}
  \hline
  Parameter & Prior & Range & Parameter & Prior & Range   \\
  \hline
  $\kappa$ & Flat & $-0.75$--$0.75$ & $\tan \beta$ & Flat & $5$--$60$ \\
  $\lambda$ & Flat & $0$--$0.75$ & $v_s/{\rm TeV}$ & Flat & $0.1$--$1.0$ \\
  $\delta$ & Flat & $-1.0$--$1.0$ & $m_N/{\rm TeV}$ & Flat & $-1.0$--$1.0$ \\
  $m_B/{\rm GeV}$ & Flat & $1$--$300$ & $m_C/{\rm GeV}$ & Flat & $1.0$--$300$ \\
  $A_t/{\rm TeV}$ & Flat & $1.0$--$3.0$ & $\mu_{\rm tot}/{\rm TeV}$ & Flat & $0.2 $--$ 1.0$ \\
  $A_\kappa/{\rm TeV}$ & Flat & $-2.0$--$2.0$ \\
  \hline
  \end{tabular}}
  \caption{Parameter space explored in this study. All input parameters used flat distributions based on their unambiguous physical interpretations. Considering the substantial radiative corrections that the third-generation squark trilinear couplings ($A_t$ and $A_b$) impose on the SM-like Higgs boson mass, we imposed
  $A_t = A_b$, where their magnitudes were treated as free variables. Non-critical SUSY-breaking parameters were fixed: $m_A = 2~{\rm TeV}$, $M_1 = 1~{\rm TeV}$,
$M_2 = 2~{\rm TeV}$, $M_3 = 3~{\rm TeV}$. All parameters were defined at the renormalization scale $Q_{input} = 1~{\rm TeV}$.}
  \label{ScanRange}

\end{table}

\subsection{\label{scan}Research strategy}
We constructed the GNMSSM model file using SARAH~\cite{Staub:2008uz,Staub:2012pb,Staub:2013tta,Staub:2015kfa}, generated particle spectra with SPheno~\cite{Porod_2012,Porod:2003um,Belanger:2014hqa,Staub:2008uz}, and used the parallelized MultiNest algorithm \cite{Feroz:2008xx} to explore the parameter space. 
In the specific parameter space scan, we set the number of live points to \( n_{\text{live}} = 8000 \) and selected the GNMSSM parameter space as shown in Table \ref{ScanRange}. The dimensionless coupling parameters $\lambda$ and $\kappa$ are mainly restricted to avoid a Landau singularity below the Grand Unified (GUT) scale \cite{Ellwanger:2009dp}, with $\lambda$ restricted to the interval [0, 0.75] and $\kappa$ to [-0.75, 0.75].
The dimensionless parameter $\delta$ was varied over the range  [-1,1] to enable a moderate cancellation between effective $\mu-$term and SUSY-breaking contributions in the matrix element ${\cal M}^2_{S, 23}$ , as defined in Eq.(\ref{M23}). 
$m_B$ and $m_C$ denote the masses of the CP-even and CP-odd singlet Higgs bosons, respectively. To obtain a light CP-even or CP-odd Higgs boson, we set the mass bounds to 
1 GeV $\leq m_B, m_C \leq$ 300 GeV.
 Due to the sizable radiative corrections from third-generation squark trilinear couplings 
$A_t$ and $A_b$ to the SM-like Higgs boson mass, we impose $A_t = A_b$, treat their magnitudes as free parameters, and vary $A_t$ in the range of 1-3 TeV.
Parameters $m_N$ and $\mu_{tot}$ are the singlino and higgsino masses, respectively. Their  scanning ranges covered a relatively broad neutralino mass spectrum. 
Parameters $A_\kappa$ and $v_s$ affect the triple Higgs couplings, and the selected scanning ranges enable a substantial  variation in the Higgs self-coupling.

To improve scanning efficiency, the likelihood function used to guide the scan  was defined as follows:
\begin{eqnarray}
    \mathcal{L} &=&  \mathcal{L}_{\rm mass} \times \mathcal{L}_{\rm HS} \times \mathcal{L}_{\rm HB}\times \mathcal{L}_{\rm B} 
   \label{Likelihood}  
\end{eqnarray}
where
\begin{itemize}
\item \textbf{Mass of the SM-like Higgs boson ($\mathcal{L}_{\text{mass}}$):} This constraint is satisfied when the mass of SM-like Higgs boson (either $h_1$ or $h_2$) lies within the range of 122 GeV to 128 GeV, accommodating theoretical and experimental uncertainties of approximately 3 GeV \cite{ATLAS:2015yey}.

\item \textbf{Higgs data fit ($\mathcal{L}_{\rm HS}$):} The properties of the SM-like Higgs boson must be consistent with the combined LHC measurements at the 95\% confidence level. 
A parameter point passes this constraint if the $p$-value computed by \textsf{HiggsSignals-2.6.2} \cite{HS2013xfa,HSConstraining2013hwa,Bechtle:2014ewa,HS2020uwn} exceeds 0.05.

\item \textbf{Extra Higgs searches ($\mathcal{L}_{\rm HB}$):} All non-SM-like Higgs bosons are required to  satisfy the 95\% confidence level exclusion limits from direct searches at LEP, Tevatron, and the LHC, as evaluated by \textsf{HiggsBounds-5.10.2} \cite{HB2008jh,HB2011sb,HBHS2012lvg,HB2013wla,HB2020pkv}. 
A parameter point is accepted only if it passes all search channels.

\item \textbf{$B$-physics observables ($\mathcal{L}_{\rm B}$):} This constraint is satisfied when the theoretical predictions for the branching ratios of $B_s \to \mu^+ \mu^-$ and $B \to X_s \gamma$ lie within the $\pm 2\sigma$ range of their respective experimental values \cite{pdg2020}.
\end{itemize}
Consequently, the likelihood $\mathcal{L} = 1$ only if all the above constraints are satisfied; otherwise, $\mathcal{L} =\exp(-100) \approx 0$. Our subsequent analysis is restricted to the parameter points with $\mathcal{L}=1$.

In addition to the above basic constraints, we also incorporate constraints from \textsf{HiggsTools} \cite{Bahl:2022igd}, which integrates the updated versions of \textsf{HiggsBounds} and \textsf{HiggsSignals}. In \textsf{HiggsTools}, \textsf{HiggsBounds} provides direct constraints from extra scalar searches, whereas \textsf{HiggsSignals} assesses the compatibility with the observed 125 GeV Higgs properties.

\subsection{\textsf{HiggsTools} constraints}

\begin{figure}[htbp]
    \centering
    \includegraphics[width=0.52\textwidth]{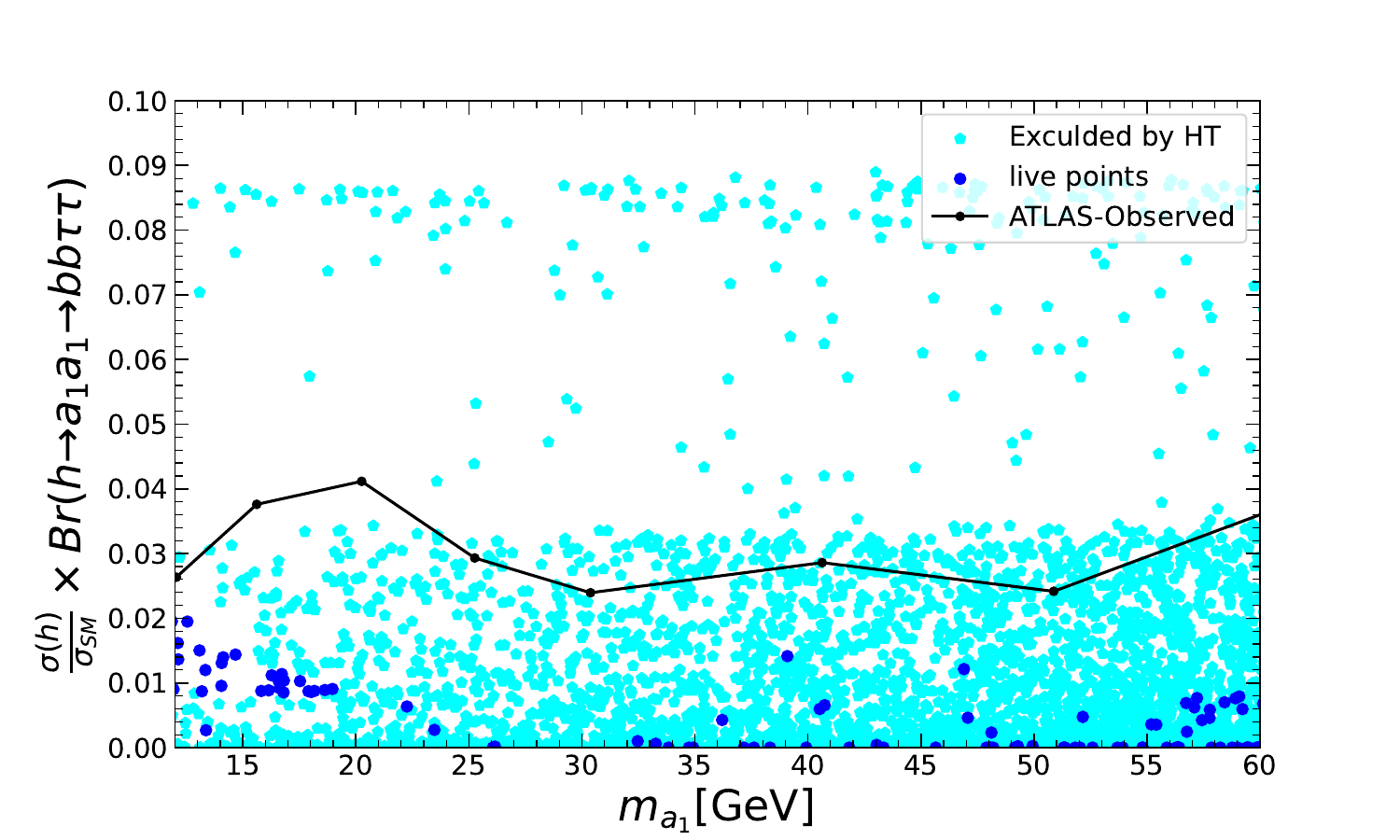} 
    \hspace{-1.cm}
    \includegraphics[width=0.52\textwidth]{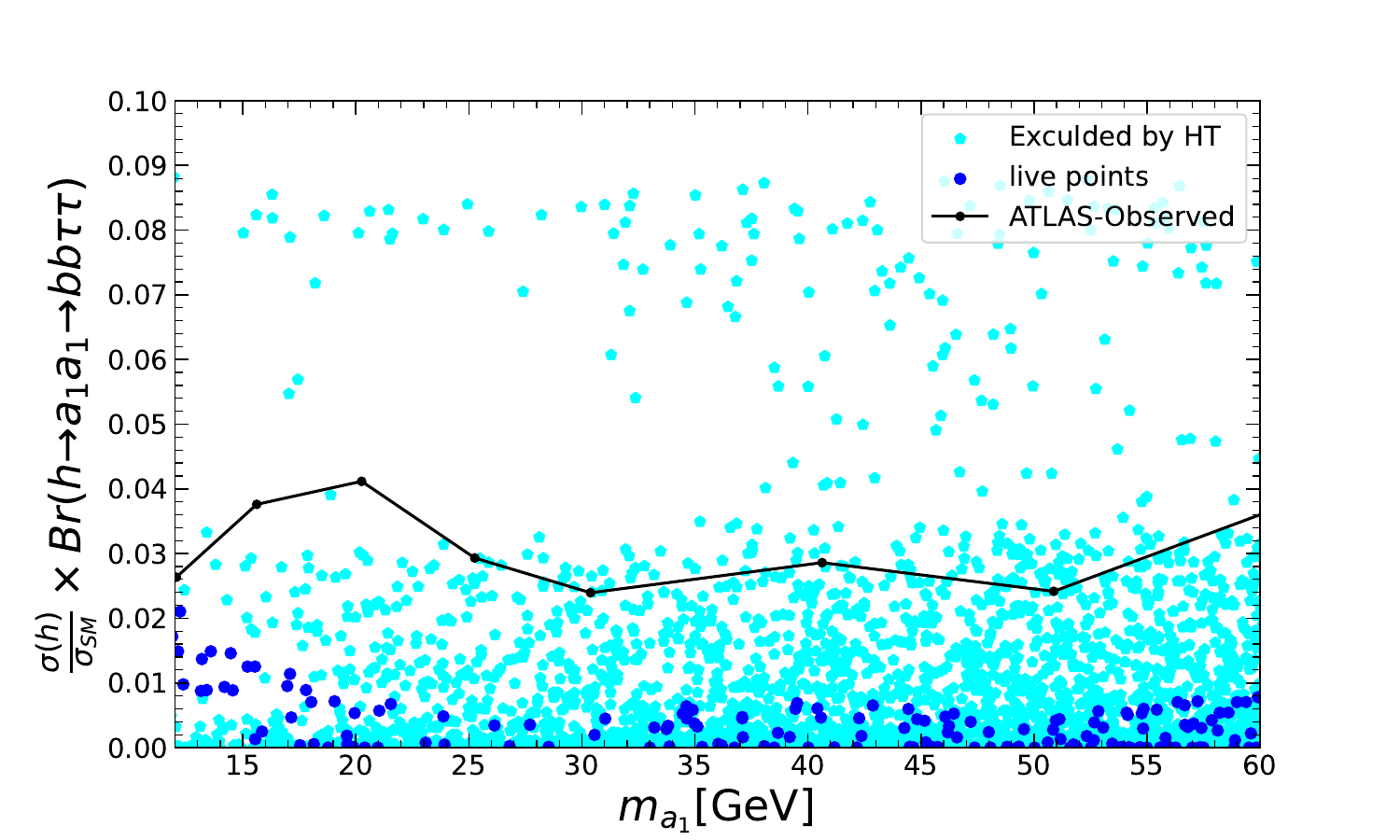} 
    \caption{Surviving parameter samples after we applied the basic constraints in Sec. 3.1 were projected onto the plane of $\sigma(h)/\sigma_{SM}(h) \times Br(h \to a_1 a_1\to bb \tau \tau)$ versus $m_{a_1}$. The blue points represent the surviving samples after further constraints by \textsf{HiggsTools} (HT); the black solid line denotes the ATLAS exclusion limit for the $h \to a_1 a_1 \to bb \tau\tau$ channel \cite{ATLAS2024}. The left and right panels correspond to the $h_1$ and $h_2$ scenarios, respectively. }
    \label{fig1}
\end{figure}

The direct search constraints for light Higgs bosons decaying into  final states such as $4b$, $\mu\mu bb$ were primarily implemented within \textsf{HiggsTools}.
To compare experimental limitations on the parameter space in the light Higgs scenario, we projected the surviving samples that satisfied the basic constraints in Sec.~3.1 onto two-dimensional planes, where the vertical axes represent $\mu_{bb\tau\tau} \equiv \sigma(h)/\sigma_{\text{SM}}(h) \times \text{Br}(h \to a_1 a_1 \to bb\tau\tau)$, as shown in Figure \ref{fig1}. The left and right panels correspond to the $h_1$ and $h_2$ scenarios, respectively. The blue points represent the surviving samples after further constraints of \textsf{HiggsTools}.
The ATLAS constraints \cite{ATLAS2024} on the decay channel $h \to a_1a_1 \to bb\tau\tau$ are applied (the solid line), and all parameter points above this limit are excluded.
Figure \ref{fig1} reveals that the upgraded \textsf{HiggsTools} imposes more stringent constraints than both the legacy \textsf{HiggsSignals} and \textsf{HiggsBounds}, as well as the  ATLAS constraints on $h \to a_1a_1 \to bb\tau\tau$.
After the constraints from \textsf{HiggsTools} had been applied, the values of $\mu_{bb\tau\tau}$ could reach 1.9\% in the $h_1$ scenario and 2.1\% in the $h_2$ scenario.

\begin{table}[tbp]
  \centering
  \vspace{0.3cm}
  \resizebox{0.7\textwidth}{!}{
  \begin{tabular}{|c|c|c|c}
  \hline
   & $h_1$ scenario & $h_2$ scenario    \\
  \hline
  live point & 143 & 201  \\
  \hline
  Excluded by both HB and HS & 964 & 1021  \\
  \hline
   Excluded only by HB  & 3287 & 2793  \\
   \hline
  Excluded only by HS  &32 & 42  \\
  \hline
  \end{tabular}}
  \caption{Numbers of surviving and excluded samples obtained with \textsf{HiggsTools}. }
  \label{excludedpoint}
\end{table}

\begin{figure}[htbp]
    \centering
    \includegraphics[width=0.52\textwidth]{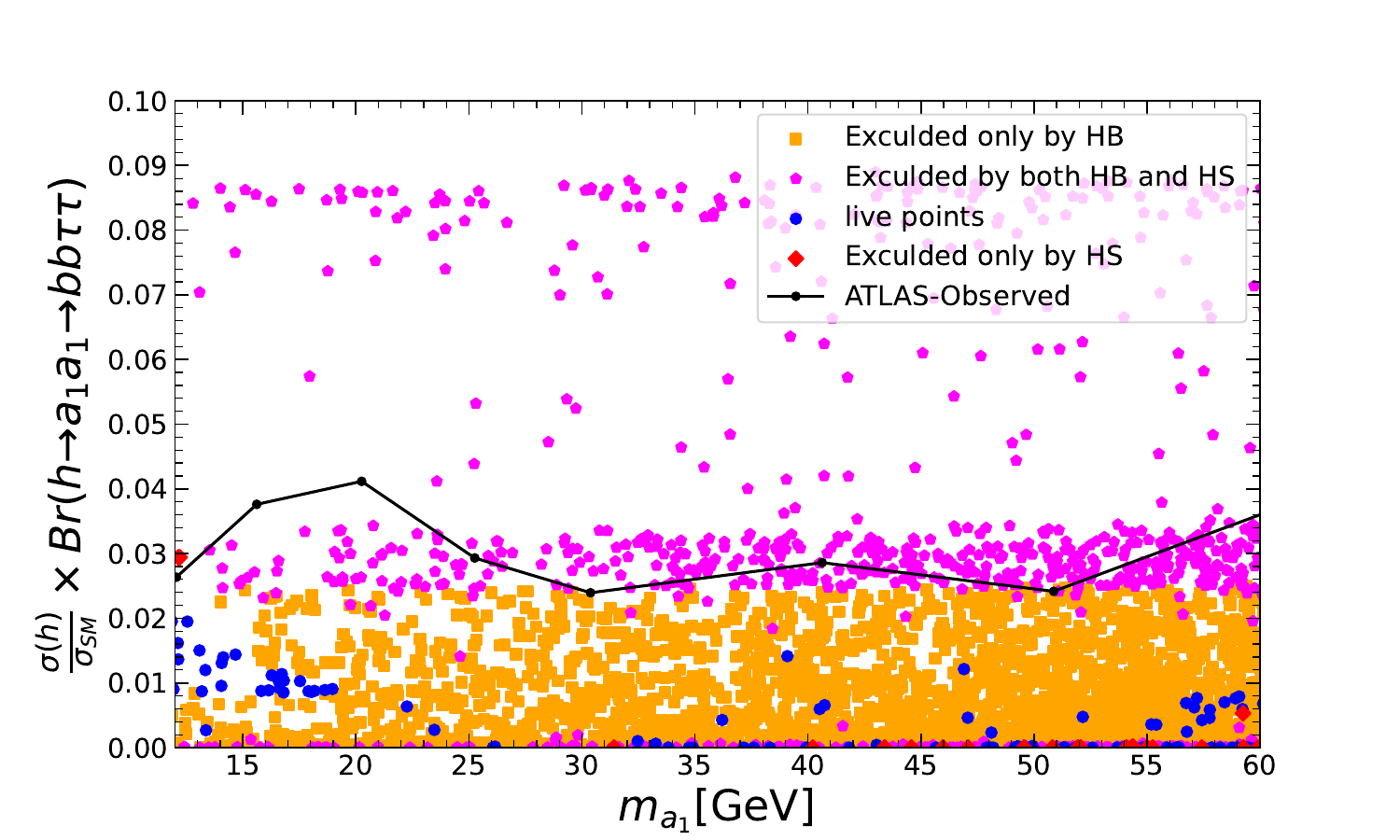} 
     \hspace{-1cm}
    \includegraphics[width=0.52\textwidth]{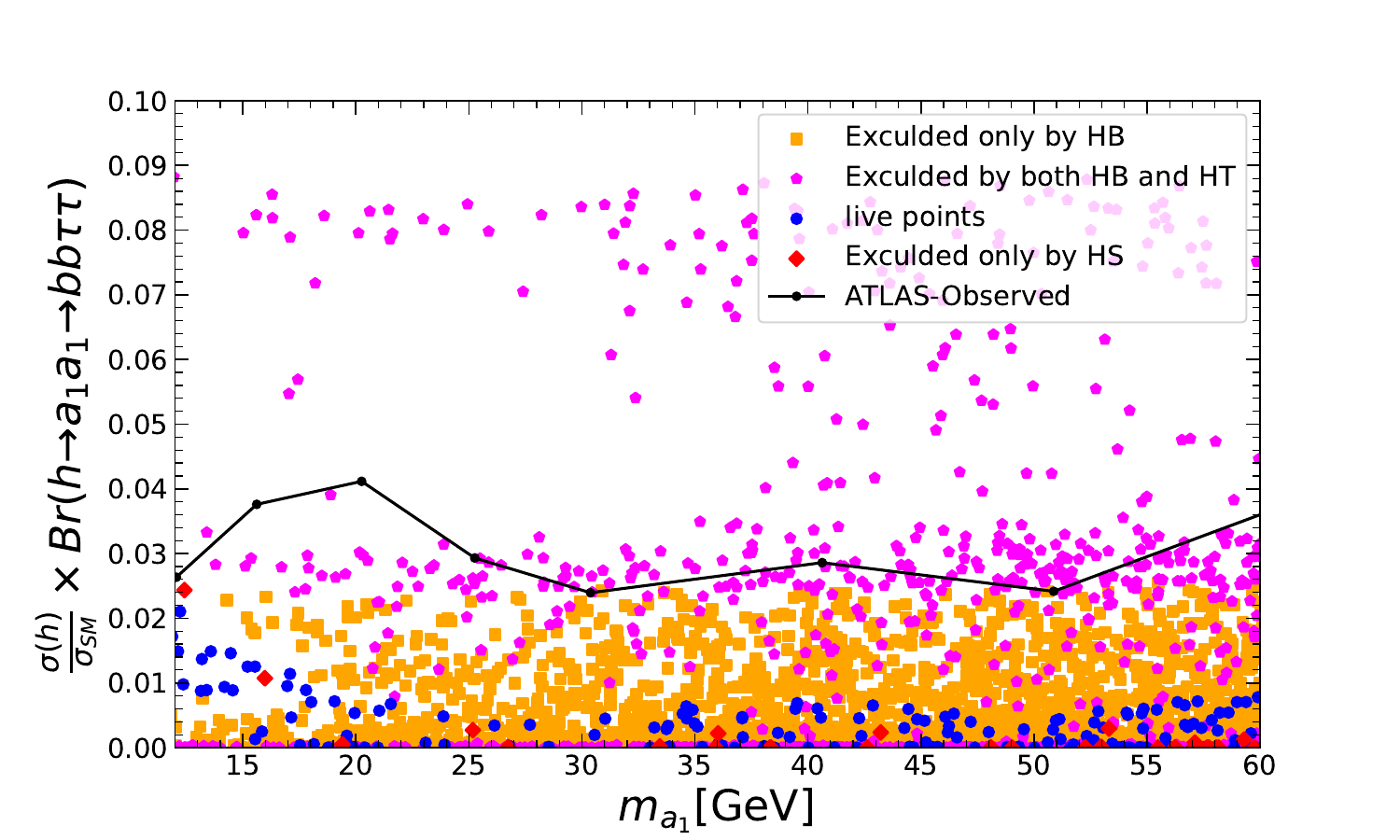} 
    \caption{Parameter samples excluded by \textsf{HiggsTools} in Figure \ref{fig1} are color-coded based on the exclusion source. The red diamond points denote samples that only \textsf{HiggsSignals} (HS) excludes, the orange square points denote samples that only \textsf{HiggsBounds} (HB) excludes, and the magenta pentagon points denote samples that both HB and HS exclude. }
    \label{fig2}
\end{figure}

To compare the constraints from \textsf{HiggsBounds} and \textsf{HiggsSignals} on the GNMSSM parameter points in the light Higgs scenario, the parameter samples excluded by \textsf{HiggsTools} in Figure \ref{fig1} are color-coded in Figure \ref{fig2} based on the exclusion source.
The red diamond markers represent samples excluded only by \textsf{HiggsSignals}, the orange square markers denote samples excluded only by \textsf{HiggsBounds}, and the magenta pentagon markers indicate samples excluded by both tools.
Figure \ref{fig2} shows that a small number of parameter points are excluded by \textsf{HiggsSignals}, whereas the vast majority of such points are also excluded by \textsf{HiggsBounds}, indicating that \textsf{HiggsBounds} has stronger exclusion power.
Table \ref{excludedpoint} shows a detailed summary of surviving and excluded samples  obtained using \textsf{HiggsTools}.
Approximately 96.7\% (94.8\%) of the parameter points were excluded by \textsf{HiggsTools} in the $h_1$ ($h_2$) scenario.

In the $h_1$ scenario, parameter points with relatively large values of $\text{Br}(h\to a_1 a_1\to\tau\tau bb)$---specifically, $\text{Br}(h\to a_1 a_1\to \tau\tau bb)\geq 2.5\%$---are excluded by both \textsf{HiggsSignals} and \textsf{HiggsBounds}. In contrast, points with smaller branching ratios are  excluded only by \textsf{HiggsBounds}. This difference arises from the fact that \textsf{HiggsBounds} imposes direct constraints, whereas \textsf{HiggsSignals} imposes indirect constraints derived from precision measurements of the 125 GeV Higgs boson properties.
Specifically,  a sufficiently large branching ratio $\text{Br}(h \to a_1 a_1)$  can affect the properties of the 125 GeV Higgs boson and implies a large $\text{Br}(h \to a_1 a_1 \to\tau\tau bb)$, although the converse is not necessarily valid. Conversely, small values of $\text{Br}(h\to a_1 a_1\to\tau\tau bb)$ always correspond to small $\text{Br}(h\to a_1 a_1)$, which has minimal impact on the 125 GeV Higgs boson properties.

\begin{figure}[htbp]
    \centering
    \includegraphics[width=0.48\textwidth]{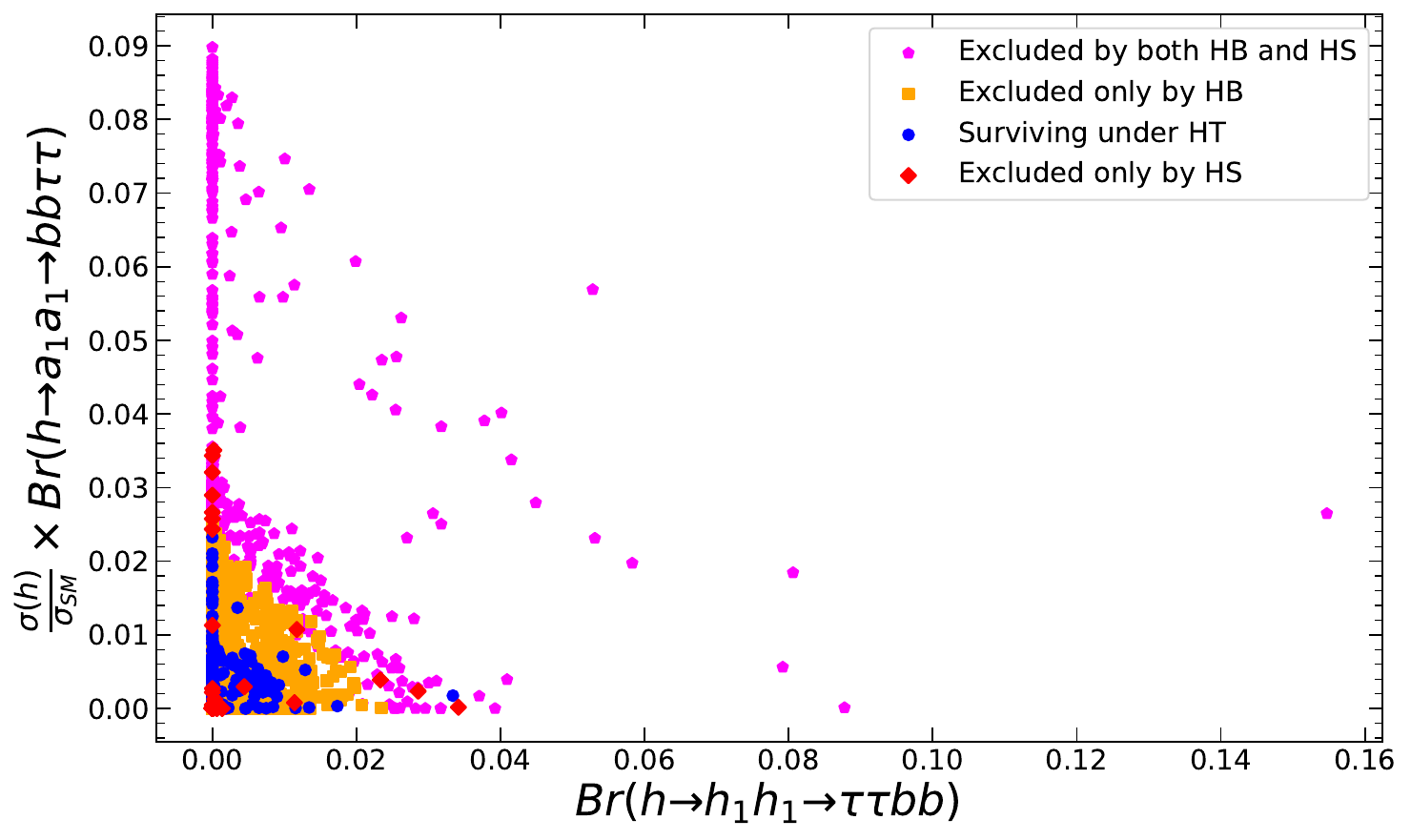} 
    \hspace{0.05cm}
    \includegraphics[width=0.48\textwidth]{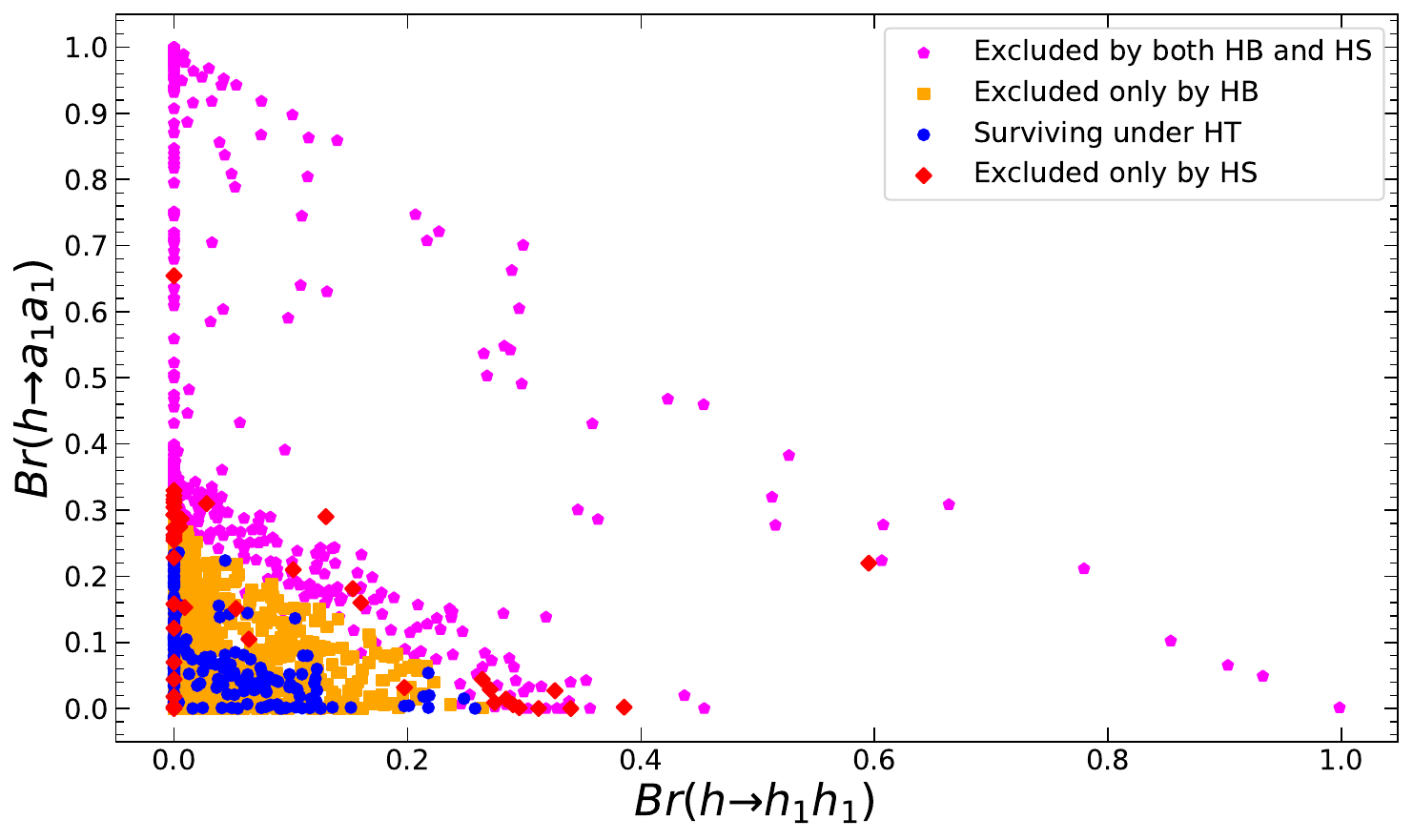} 
    \caption{Following the right panel of Figure \ref{fig2} in the $h_2$ scenario, the left panel here plots $\sigma(h)/\sigma_{SM}(h) \times \text{Br}(h \to a_1 a_1\to bb \tau \tau)$ against
    $\text{Br}(h \to h_1h_1\to \tau\tau bb)$, and the right panel displays the relationship between Br$(h \to a_1 a_1)$ and Br$(h \to h_1 h_1)$.}
    \label{fig3}
\end{figure}

The $h_2$ scenario shows characteristics similar to the $h_1$ scenario, but with distinct features. As shown in the right panel of Figure \ref{fig2}, 
some parameter points with $Br(h\to a_1 a_1\to \tau\tau bb)\leq 2.5\%$ (indicated by magenta points) are excluded by \textsf{HiggsSignals}, since the exotic decay channel $h \to h_1h_1$ becomes kinematically accessible in this scenario with an enhanced branching fraction. This enhancement may affect the properties of the SM-like Higgs boson and the constraints that \textsf{HiggsSignals} imposes. The left panel of Figure \ref{fig3} shows the relationship between $Br(h\to a_1 a_1\to \tau\tau bb)$ and $Br(h \to h_1h_1\to \tau\tau bb)$, and the right panel displays Br$(h \to a_1 a_1)$ versus Br$(h \to h_1 h_1)$ in the $h_2$ scenario. It reveals an anti-correlation between these branching ratios across the parameter space, which validates the proposed explanation. The right panel also indicates that the points featuring a relatively large $Br(h \to h_1 h_1)$ are predominantly excluded by the \textsf{HiggsSignals}.

\begin{figure}[htbp]
    \centering
    \includegraphics[width=0.49\textwidth]{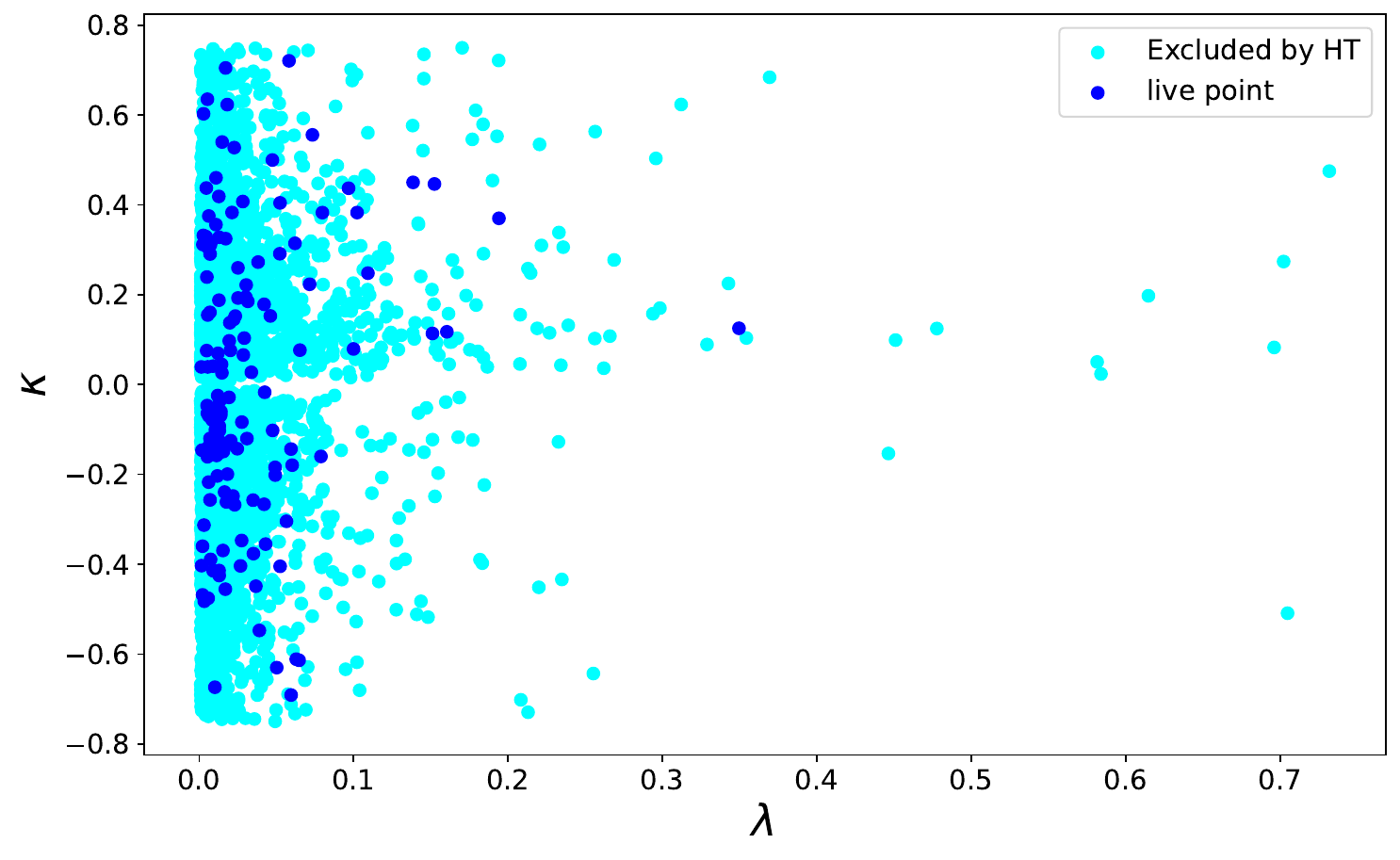} 
    \hfill
    \includegraphics[width=0.49\textwidth]{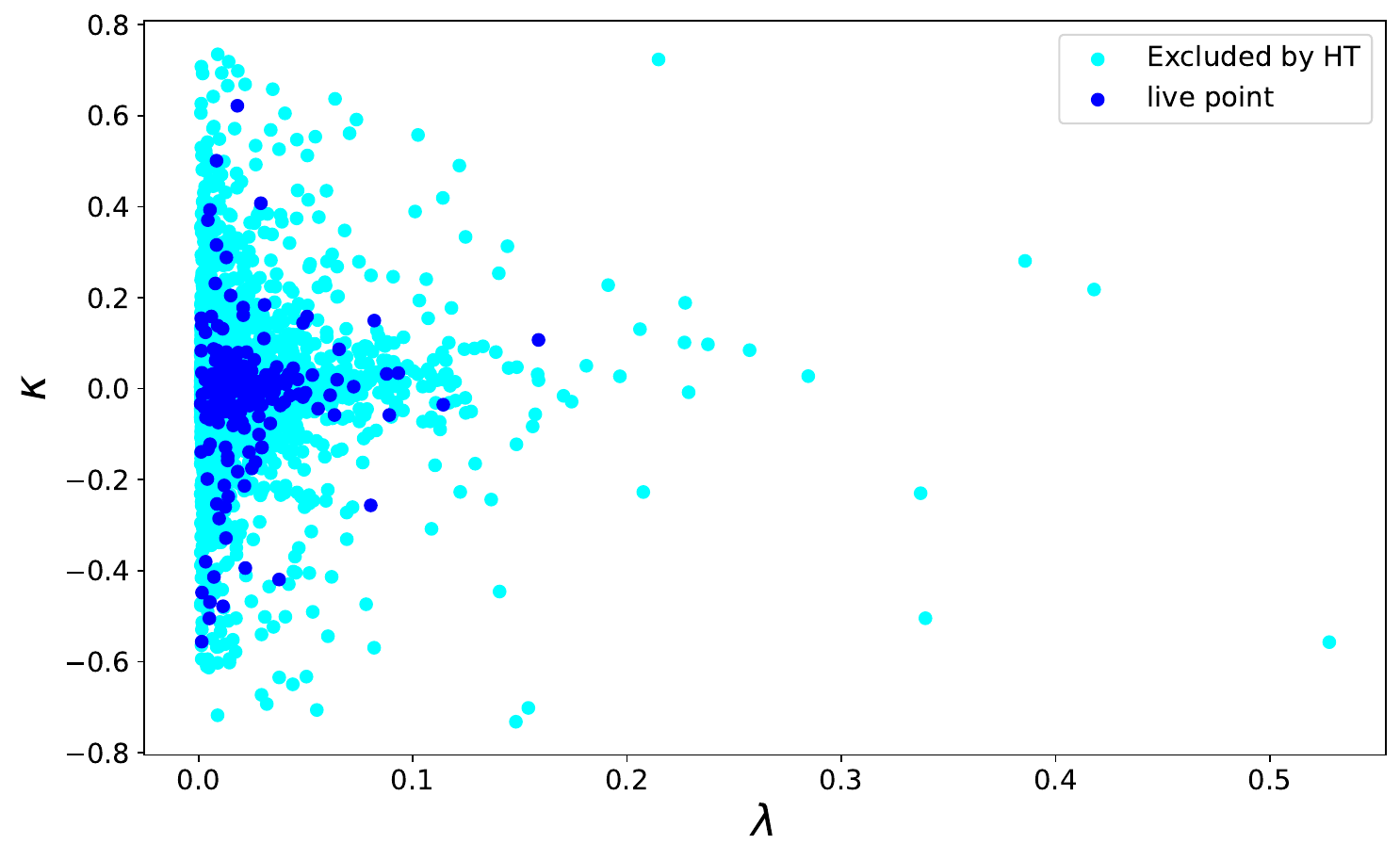} 
    \caption{Relationship between parameters $\lambda$ and $\kappa$, similar to Figure \ref{fig1}.}
    \label{fig4}
\end{figure}

\begin{figure}[htbp]
    \centering
    \includegraphics[width=0.7\textwidth]{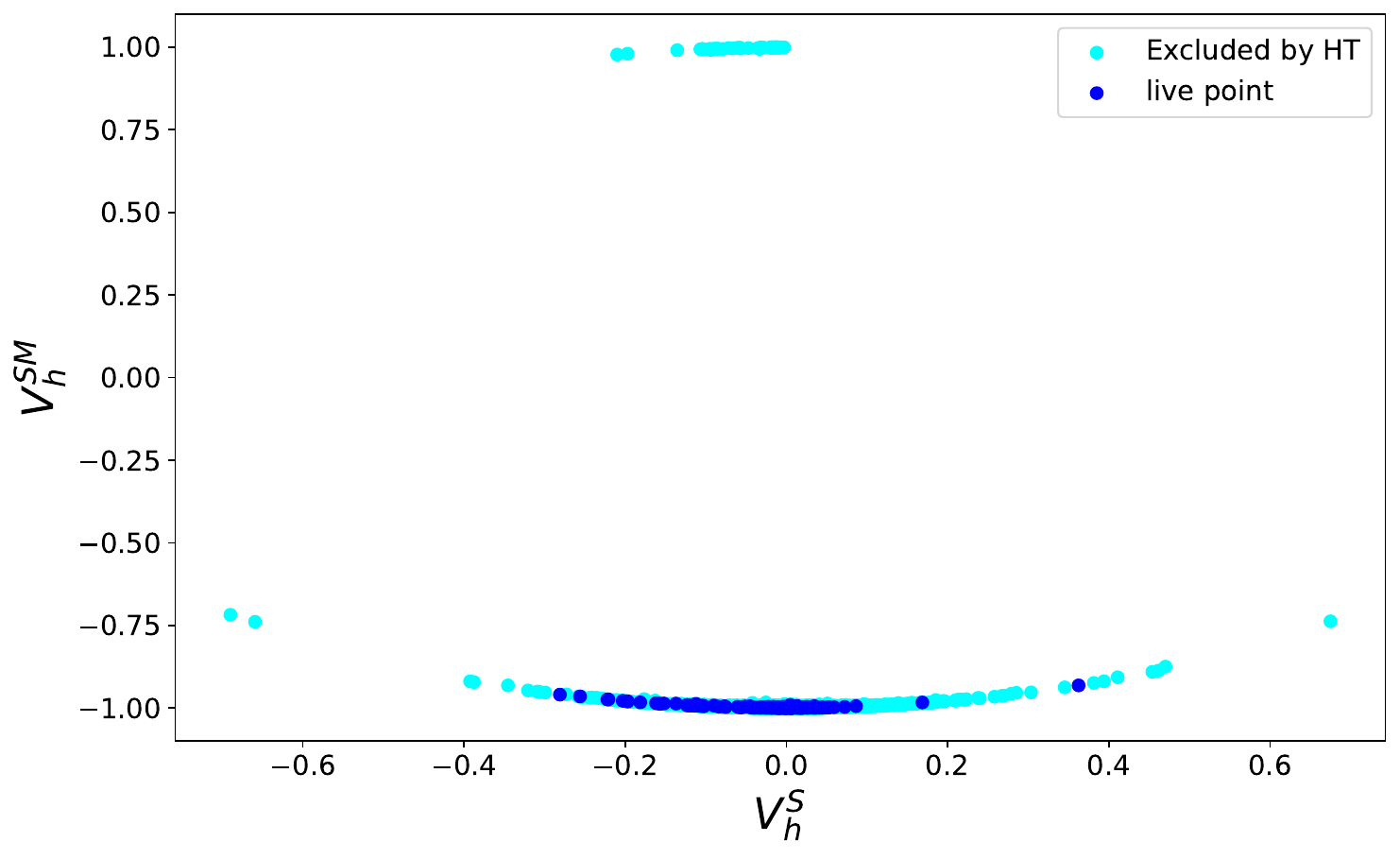} 
    \caption{SM-like Higgs boson component in the $h_1$ scenario, similar to Figure ~\ref{fig1}.}
    \label{fig5}
\end{figure}

\begin{figure}[htbp]
    \centering
    \includegraphics[width=0.49\textwidth]{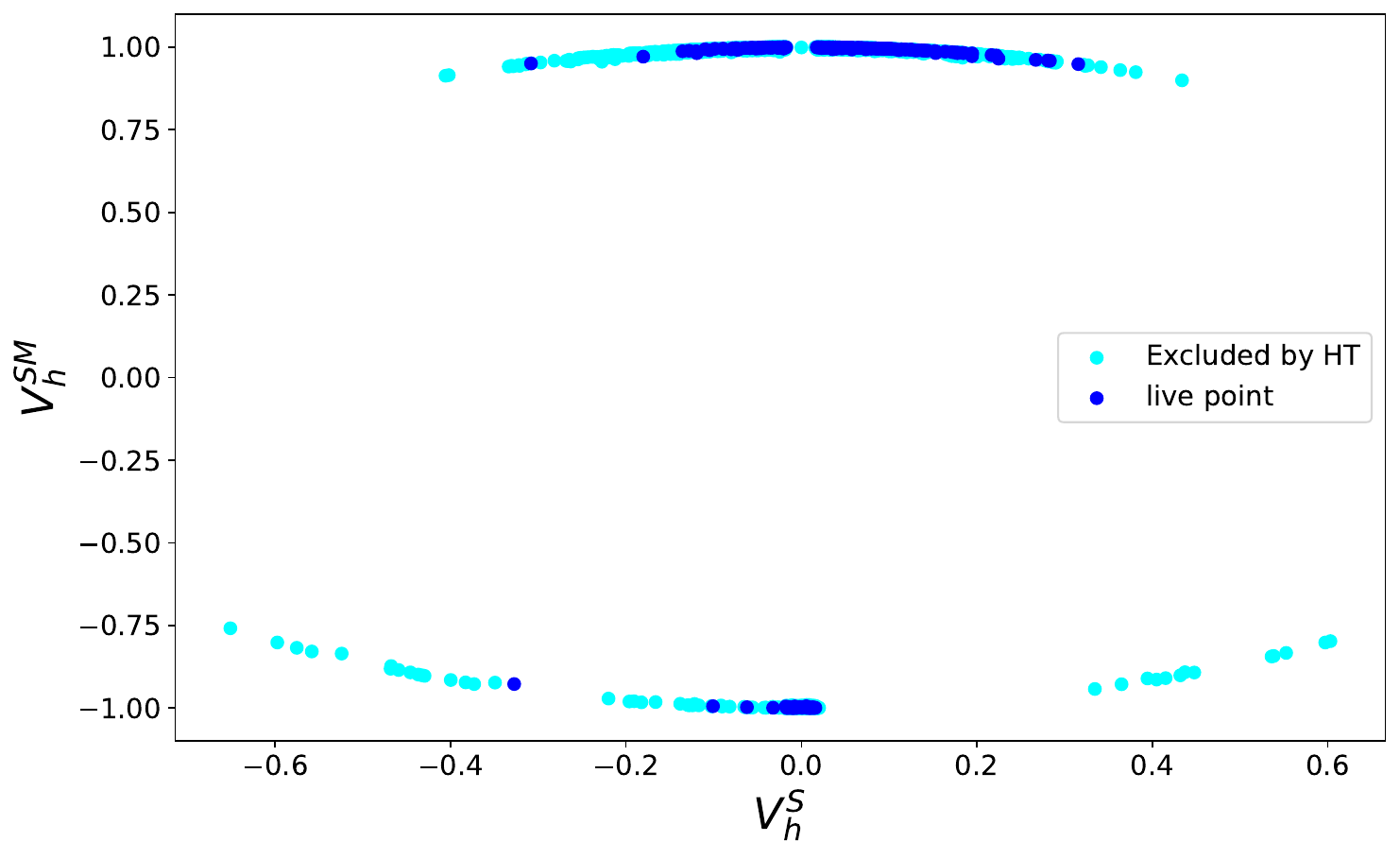} 
    \hfill
    \includegraphics[width=0.49\textwidth]{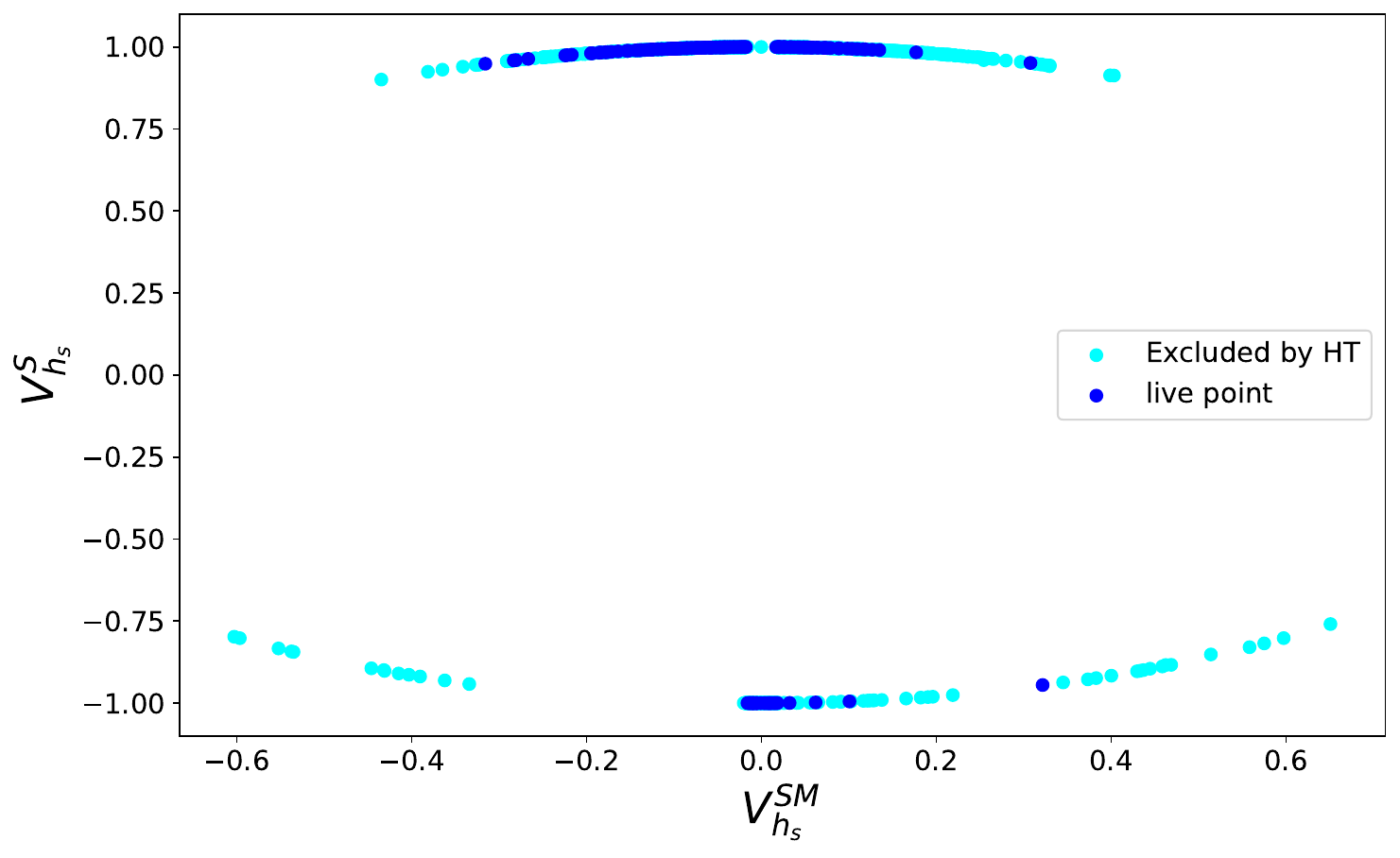} 
    \caption{$h_2$ component (left panel) and $h_1$ component (right panel) in the $h_2$ scenario, similar to Figure ~\ref{fig1}.}
    \label{fig6}
\end{figure}

The branching ratios for the exotic decays of $h \to a_1 a_1$ and $h \to h_1 h_1$ depend on the corresponding  triple Higgs couplings $C_{h a_1 a_1}$ and $C_{h h_1 h_1}$, as defined in Eqs. (\ref{haa}) and (\ref{hhh}). These couplings are determined primarily by parameters $\lambda$ and $\kappa$, along with the SM-like fraction $V_h^{\rm SM}$ of the $h$ boson. In the $h_2$ scenario, $C_{h h_1 h_1}$ is also modulated by the singlet fraction $V_{h_s}^{\rm S}$ of the singlet-dominated state $h_s$. Figure \ref{fig4} displays the correlation between $\lambda$ and $\kappa$ under \textsf{HiggsTools} constraints, which indicates that $\lambda \leq 0.1$ for most viable parameter points, which is consistent with experimental preferences for minimal doublet-singlet mixing. 
However, when $\lambda$ is too small, the $\sim\lambda^2$ term in Eq. (\ref{h-mass}) has negligible contribution, and the calculation of the SM-like Higgs boson mass essentially reduces to that in the MSSM. 
The level of fine-tuning in this regime was comparable to that in the MSSM with no significant improvement.
For benchmark point 4 
 in Table \ref{Tab3}, with $m_{\tilde t _1} = 1952.2\ \text{GeV}$ and $m_{\tilde t _2} = 2071.5\ \text{GeV}$, we found that achieving $m_h \simeq 125\ \text{GeV}$ resulted in $\Delta^2 /m_h^2 \simeq 92\%$, which indicates a high degree of fine-tuning.

Figures \ref{fig5} and \ref{fig6} reveal the Higgs boson compositions. Both $h_1$ and $h_2$ scenarios require that the observed $h$ boson is predominantly SM-like ($V_h^{\rm SM} \geq 0.93$) with suppressed singlet admixture ($V_h^{\rm S} \leq 0.32$). For the $h_2$ scenario, the lighter $h_s$ state must exhibit high singlet purity ($V_{h_s}^{\rm S} \geq 0.94$) and negligible SM-like components ($V_{h_s}^{\rm SM} \leq 0.32$).
Figure \ref{fig6} further illustrates that $V_h^S \simeq - V_{h_s}^{\rm SM}$, confirming the theoretical derivation presented in Eq.(\ref{Approximations}).

\begin{table}[htbp]
  \centering
  \resizebox{\textwidth}{!}{
  \begin{tabular}{lr@{}ll|lr@{}ll}
      \toprule
      \multicolumn{4}{c|}{\textbf{Point P1}} & \multicolumn{4}{c}{\textbf{Point P2}} \\ \midrule
      Parameter     & &  & Value  & Parameter & &  & Value  \\ \midrule
    $\lambda$, $\kappa$,  & & & $0.015,-0.369$ & $\lambda$, $\kappa$ & & & $0.017,-0.455$ \\
    $\tan\beta$, $\delta$   & & & $41.5,0.02$ & $\tan\beta$, $\delta$ & & & $37.6,0.08$ \\
      $v_s$,  $\mu_{\rm tot}$, $m_N$ & & & $212.8,468.9,23.6$ & $v_s$,  $\mu_{\rm tot}$, $m_N$ & & & $811.7,342.0,147.7$  \\
      $m_B$, $m_C$ & & & $208.8,144.6$ & $m_B$,$m_C$ & & & $181.6,23.3$ \\
      $A_t$, $A_\kappa$ & & & $2536.3,-685.3$ & $A_t$,$A_\kappa$ & & & $2592.3,-1152.8$ \\
      \midrule

      Particle      & & & Mass Spectrum & Particle & & & Mass Spectrum \\
      $\tilde{\chi}_1^0$,$\tilde{\chi}_2^0$, $\tilde{\chi}_3^0$ & & & $23.5,478.2,483.5$ & $\tilde{\chi}_1^0$,$\tilde{\chi}_2^0$, $\tilde{\chi}_3^0$ & & & $146.5,349.2,354.2$ \\
      $\tilde{\chi}_4^0$,  $\tilde{\chi}_5^0$ & & & $1008.3,2020.8$ & $\tilde{\chi}_4^0$, $\tilde{\chi}_5^0$ & & & $1008.1,2020.8$  \\
      $\tilde{\chi}_1^{\pm}$,  $\tilde{\chi}_2^{\pm}$ & & & $481.4,2021.1$ & $\tilde{\chi}_1^{\pm}$, $\tilde{\chi}_2^{\pm}$ & & & $352.2,2021.0$  \\
      $h_s$, $h$, $H$  & & & $154.8,125.1,2414.2$ & $h_s$, $h$, $H$  & & & $188.9,125.3,2273.6$ \\
      $a_s$, $A_H$ & & &  $48.0, 2414.1$ &  $a_s$, $A_H$ & & &  $16.6, 2273.6$  \\        \midrule
      Primary Decays  & & & Branching Ratios [\%]  & Primary Decays  & & & Branching Ratios [\%] \\
      $h\to b \bar b$  & & & $47.5$  & $h\to b \bar b$  & & & $49.4$ \\
      $h\to W^+W^-$  & & & $22.4$  & $h\to W^+W^-$  & & & $24.0$ \\
      $h\to a_s a_s$  & & & $16.1$  & $h\to a_s a_s$  & & & $11.9$ \\  \midrule

     Rotation Matrix     & & & Element Value & Rotation Matrix & & & Element Value \\
      \multicolumn{2}{l}{$V_{h_s}^S$,$V_{h_s}^{SM}$,$V_{h}^S$,$V_{h}^{SM}$} & \multicolumn{2}{l|}{$-0.99996,\ 0.0005,\ -0.0003,\ -0.9997$} &
      \multicolumn{2}{l}{$V_{h_s}^S$,$V_{h_s}^{SM}$,$V_{h}^S$,$V_{h}^{SM}$} & \multicolumn{2}{l}{$-0.99996,\ -0.005,\ 0.005,\ -0.9996$} \\

      \multicolumn{2}{l}{$N_{11}$,$N_{12}$,$N_{13}$,$N_{14}$,$N_{15}$} & \multicolumn{2}{l|}{$0.0003,-0.0002,0.0002,-0.0055,0.99998$} &
      \multicolumn{2}{l}{$N_{11}$,$N_{12}$,$N_{13}$,$N_{14}$,$N_{15}$} & \multicolumn{2}{l}{$-0.00059,\ 0.0004,\ -0.004,\ 0.00997,\ -0.9999$} \\

      \multicolumn{2}{l}{$N_{21}$,$N_{22}$,$N_{23}$,$N_{24}$,$N_{25}$} & \multicolumn{2}{l|}{-0.0599,\ 0.036,\ -0.71,\ 0.70,\ 0.004} &
      \multicolumn{2}{l}{$N_{21}$,$N_{22}$,$N_{23}$,$N_{24}$,$N_{25}$} & \multicolumn{2}{l}{0.048,\ -0.033,\ 0.71,\ -0.70,\ -0.00996} \\

      \multicolumn{2}{l}{$N_{31}$,$N_{32}$,$N_{33}$,$N_{34}$,$N_{35}$} & \multicolumn{2}{l|}{0.02,\ -0.02,\ -0.71,\ -0.71,\ -0.0037} &
      \multicolumn{2}{l}{$N_{31}$,$N_{32}$,$N_{33}$,$N_{34}$,$N_{35}$} & \multicolumn{2}{l}{-0.022,\ 0.023,\ 0.70,\ 0.71,\ 0.004} \\

      \multicolumn{2}{l}{$N_{41}$,$N_{42}$,$N_{43}$,$N_{44}$,$N_{45}$} & \multicolumn{2}{l|}{0.998,\ 0.0045,\ -0.028,\ 0.056,\ 0} &
      \multicolumn{2}{l}{$N_{41}$,$N_{42}$,$N_{43}$,$N_{44}$,$N_{45}$} & \multicolumn{2}{l}{-0.998,\ -0.004,\ 0.02,\ -0.05,\ 0} \\

      \multicolumn{2}{l}{$N_{51}$,$N_{52}$,$N_{53}$,$N_{54}$,$N_{55}$} & \multicolumn{2}{l|}{0.0019,\ -0.999,\ -0.01,\ 0.04,\ 0} &
      \multicolumn{2}{l}{$N_{51}$,$N_{52}$,$N_{53}$,$N_{54}$,$N_{55}$} & \multicolumn{2}{l}{0.002,\ -0.999,\ -0.0078,\ 0.04,\ 0} \\ \midrule

      Primary Annihilation & & & Fraction [\%]               & Primary Annihilation & & & Fraction [\%] \\
     $\tilde{\chi}_1^0  \tilde{\chi}_1^0\to b \bar{b} $ & & &~~$79.97$ & $\tilde{\chi}_1^0 \tilde{\chi}_1^{0} \to h_s a_s $ & & &~~$81.2$ \\
     $\tilde{\chi}_1^0 \tilde{\chi}_1^{0} \to \tau^+ \tau^- $ & & &~~$19.74$ & $\tilde{\chi}_1^0 \tilde{\chi}_1^0 \to a_s a_s$ & & &~~$18.79$
      \\ \midrule

      DM Observable     & & & Value & DM Observable & & & Value \\
      $\Omega h^2$ & & & $0.000086$ & $\Omega h^2$ & & & $0.001593$ \\
      $\sigma^{\rm SI}_{p,n}/(10^{-49} \text{cm}^2)$ & & & $6.8,7.9 $ &
      $\sigma^{\rm SI}_{p,n}/(10^{-48} \text{cm}^2)$ & & & $4.1,4.8$ \\
      $\sigma^{\rm SD}_{p,n}/(10^{-50} \text{cm}^2)$ & & & $2.5,1.9 $ &
      $\sigma^{\rm SD}_{p,n}/(10^{-48} \text{cm}^2)$ & & & $3.6,2.8$ \\    \bottomrule
  \end{tabular}
  }
  \caption{Benchmark points P1 and P2 compatible with all experimental constraints. Both points correspond to singlino-dominated DM in the $h_1$ scenario. All mass parameters are given in GeV. \label{Tab2}}
\end{table}

\begin{table}[htbp]
  \centering
  \resizebox{\textwidth}{!}{
  \begin{tabular}{lr@{}ll|lr@{}ll}
      \toprule
      \multicolumn{4}{c|}{\textbf{Point P3}} & \multicolumn{4}{c}{\textbf{Point P4}} \\ \midrule
      Parameter     & &  & Value  & Parameter & &  & Value  \\ \midrule
    $\lambda$, $\kappa$,  & & & $0.16,0.117$ & $\lambda$, $\kappa$ & & & $0.026,-0.016$ \\
    $\tan\beta$, $\delta$   & & & $20.4,-0.02$ & $\tan\beta$, $\delta$ & & & $23.7,0.108$ \\
      $v_s$,  $\mu_{\rm tot}$, $m_N$ & & & $739.4,269.5,776.5$ &
      $v_s$,  $\mu_{\rm tot}$, $m_N$ & & & $872.6,405.6,-655.4$  \\
      $m_B$, $m_C$ & & & $282.3,240.0$ & $m_B$,$m_C$ & & & $51.2,73.5$ \\
      $A_t$, $A_\kappa$ & & & $2694.0,-1683.3$ &
      $A_t$,$A_\kappa$ & & & $2720.2,1859.9$ \\
      \midrule

      Particle      & & & Mass Spectrum & Particle  & & & Mass Spectrum \\
      $\tilde{\chi}_1^0$,$\tilde{\chi}_2^0$, $\tilde{\chi}_3^0$ & & & $273.8,279.6,777.0$ &
      $\tilde{\chi}_1^0$,$\tilde{\chi}_2^0$, $\tilde{\chi}_3^0$ & & & $413.0,418.1,655.4$ \\
      $\tilde{\chi}_4^0$,  $\tilde{\chi}_5^0$ & & & $1008.1,2020.8$ & $\tilde{\chi}_4^0$, $\tilde{\chi}_5^0$ & & & $1008.2,2020.7$  \\
      $\tilde{\chi}_1^{\pm}$,  $\tilde{\chi}_2^{\pm}$ & & & $277.3,2021.0$ & $\tilde{\chi}_1^{\pm}$, $\tilde{\chi}_2^{\pm}$ & & & $416.1,2021.0$  \\
      $h_s$, $h$, $H$  & & & $169.1,125.3,2104.7$ &
      $h_s$, $h$, $H$  & & & $21.77,125.4,2177.9$ \\
      $a_s$, $A_H$ & & &  $61.2, 2103.9$ &  $a_s$, $A_H$ & & &  $56.8, 2177.9$  \\
      \midrule

      Primary Decays  & & & Branching Ratios [\%]  & Primary Decays  & & & Branching Ratios [\%] \\
      $h\to b \bar b$  & & & $52.1$  & $h\to b \bar b$  & & & $48.4$ \\
      $h\to W^+W^-$  & & & $24.0$  & $h\to W^+W^-$  & & & $23.7$ \\
      $h\to a_s a_s /h_s h_s$  & & & $8.8/0$  & $h\to a_s a_s/h_s h_s$  & & & $3.6/9.96$ \\
       \midrule

      Rotation Matrix     & & & Element Value & Rotation Matrix & & & Element Value \\
      \multicolumn{2}{l}{$V_{h_s}^S$,$V_{h_s}^{SM}$,$V_{h}^S$,$V_{h}^{SM}$} & \multicolumn{2}{l|}{$-0.999,\ 0.03,\ -0.03,\ -0.998$} &
      \multicolumn{2}{l}{$V_{h_s}^S$,$V_{h_s}^{SM}$,$V_{h}^S$,$V_{h}^{SM}$} & \multicolumn{2}{l}{$0.9997,\ -0.021,\ 0.021,\-0.9989$} \\

      \multicolumn{2}{l}{$N_{11}$,$N_{12}$,$N_{13}$,$N_{14}$,$N_{15}$} & \multicolumn{2}{l|}{$0.044,-0.033, 0.71,-0.70,0.037$} &
      \multicolumn{2}{l}{$N_{11}$,$N_{12}$,$N_{13}$,$N_{14}$,$N_{15}$} & \multicolumn{2}{l}{$0.054,\ -0.035,\ 0.71,\ -0.70,\ -0.0029$} \\

      \multicolumn{2}{l}{$N_{21}$,$N_{22}$,$N_{23}$,$N_{24}$,$N_{25}$} & \multicolumn{2}{l|}{$-0.023,\ 0.023,\ 0.70,\ 0.71,\ 0.019$} &
      \multicolumn{2}{l}{$N_{21}$,$N_{22}$,$N_{23}$,$N_{24}$,$N_{25}$} & \multicolumn{2}{l}{$0.021,\ -0.022,\ -0.71,\ -0.71,\ 0.014$} \\

      \multicolumn{2}{l}{$N_{31}$,$N_{32}$,$N_{33}$,$N_{34}$,$N_{35}$} & \multicolumn{2}{l|}{$-0.0001,\ 0.0009,\ -0.04,\ 0.01,\ 0.999$} &
      \multicolumn{2}{l}{$N_{31}$,$N_{32}$,$N_{33}$,$N_{34}$,$N_{35}$} & \multicolumn{2}{l}{$-0.0001,\ 0.0002,\ 0.012,\ 0.0078,\ 0.999899$} \\

      \multicolumn{2}{l}{$N_{41}$,$N_{42}$,$N_{43}$,$N_{44}$,$N_{45}$} & \multicolumn{2}{l|}{$-0.9987,\ -0.004,\ 0.015,\ -0.047,\ 0.001$} &
      \multicolumn{2}{l}{$N_{41}$,$N_{42}$,$N_{43}$,$N_{44}$,$N_{45}$} & \multicolumn{2}{l}{$-0.998,\ -0.004,\ 0.02,\ -0.05,\ 0$} \\

      \multicolumn{2}{l}{$N_{51}$,$N_{52}$,$N_{53}$,$N_{54}$,$N_{55}$} & \multicolumn{2}{l|}{$-0.002,\ 0.999,\ 0.007,\ -0.039,\ -0.0001$} &
      \multicolumn{2}{l}{$N_{51}$,$N_{52}$,$N_{53}$,$N_{54}$,$N_{55}$} & \multicolumn{2}{l}{$0.002,\ -0.999,\ -0.0098,\ 0.04,\ 0$} \\ \midrule

      Primary Annihilation & & & Fraction [\%]     & Primary Annihilation & & & Fraction [\%] \\
     $\tilde{\chi}_1^0  \tilde{\chi}_1^{\pm}\to d \bar{u} / \bar{d} u$ & & &~~$10.58$ & $\tilde{\chi}_1^0 \tilde{\chi}_1^{\pm} \to d \bar{u} / \bar{d} u$ & & &~~$10.05$ \\
     $\tilde{\chi}_1^0 \tilde{\chi}_1^{\pm} \to s \bar{c} / \bar{s} c $ & & &~~$10.58$ & $\tilde{\chi}_1^0 \tilde{\chi}_1^{\pm} \to s \bar{c} / \bar{s} c$ & & &~~$10.05$ \\
     $\tilde{\chi}_1^0  \tilde{\chi}_1^{\pm} \to  b \bar{t} / t \bar{b} $ & & &~~$7.8$& $\tilde{\chi}_1^0 \tilde{\chi}_1^{\pm} \to b \bar{t} / t \bar{b} $ & & &~~$6.66$ \\
     $\tilde{\chi}_1^0 \tilde{\chi}_1^{0} \to W^+W^- $ & & &~~$6.7$ &
     $\tilde{\chi}_1^0 \tilde{\chi}_1^{0} \to W^+W^- $ & & &~~$4.84$ \\
     $\tilde{\chi}_1^0 \tilde{\chi}_1^{0} \to  ZZ$ & & &~~$5.21$ &
     $\tilde{\chi}_2^0 \tilde{\chi}_1^{\pm} \to d \bar{u} / \bar{d} u $ & & &~~$4.41$ \\
     $\tilde{\chi}_1^0 \tilde{\chi}_1^{\pm} \to  e \bar{\nu_e} / \bar{\nu_e} e$ & & &~~$3.53$ &
     $\tilde{\chi}_2^0 \tilde{\chi}_1^{\pm} \to s \bar{c} / \bar{s} c $ & & &~~$4.41$ \\
     $\tilde{\chi}_1^0 \tilde{\chi}_1^{\pm} \to  \mu \bar{\nu_{\mu}} / \bar{\nu_{\mu}} \mu$ & & &~~$3.53$     &
     $\tilde{\chi}_1^0 \tilde{\chi}_1^{0} \to  ZZ$ & & &~~$3.81$ \\
     $\tilde{\chi}_1^0 \tilde{\chi}_1^{\pm} \to  \tau \bar{\nu_{\tau}} / \bar{\nu_{\tau}} \mu$ & & &~~$3.50$  &
     $\tilde{\chi}_1^0 \tilde{\chi}_1^{\pm} \to  e \bar{\nu_e} / \bar{\nu_e} e $ & & &~~$3.35$ \\
     $\tilde{\chi}_2^0 \tilde{\chi}_1^{\pm} \to d \bar{u} / \bar{d} u $ & & &~~$3.17$  &
     $\tilde{\chi}_1^0 \tilde{\chi}_1^{\pm} \to   \mu \bar{\nu_{\mu}} / \bar{\nu_{\mu}} \mu $ & & &~~$3.35$ \\
     $\tilde{\chi}_2^0 \tilde{\chi}_1^{\pm} \to s \bar{c} / \bar{s} c $ & & &~~$3.17$ &
     $\tilde{\chi}_1^0 \tilde{\chi}_1^{\pm} \to   \tau \bar{\nu_{\tau}} / \bar{\nu_{\tau}} \tau $ & & &~~$3.27$ \\
   \midrule

      DM Observable     & & & Value & DM Observable & & & Value \\
      $\Omega h^2$ & & & $0.0091$ & $\Omega h^2$ & & & $0.02$ \\
      $\sigma^{\rm SI}_{p,n}/(10^{-48} \text{cm}^2)$ & & & $6.32,5.47$ &
      $\sigma^{\rm SI}_{p,n}/(10^{-48} \text{cm}^2)$ & & & $1.78,4.85$ \\
      $\sigma^{\rm SD}_{p,n}/(10^{-43} \text{cm}^2)$ & & & $1.64,1.26$ &
      $\sigma^{\rm SD}_{p,n}/(10^{-43} \text{cm}^2)$ & & & $4.08,3.13$ \\   \bottomrule
     \end{tabular}
  }
  \caption{Benchmark points P3 and P4 that satisfy all experimental constraints. Points P3 and P4 represent higgsino-dominated DM in the $h_1$ and $h_2$ scenarios, respectively. All mass parameters are given in GeV.  \label{Tab3}}
\end{table}

\subsection{Dark matter physics}
To further elucidate the DM phenomenology in the light Higgs scenario of the GNMSSM parameter space, the package MicrOMEGAs ~\cite{Belanger:2001fz, Belanger:2005kh,Belanger:2006is,Belanger:2010pz,Belanger:2013oya,Barducci:2016pcb,Belanger:2018mqt} was used to calculate the DM relic density and the spin-dependent (SD) and spin-independent (SI) DM-nucleon cross sections.
It was assumed that a large DM population existed in the early universe and froze out to match the current Planck observation, $\Omega_{\rm DM} h^2=0.12 \pm 0.01$ \cite{Planck:2018vyg}.
In this study, the relic density of DM (the lightest neutralino $\tilde \chi_1^0$) was required to be below the central value of 0.12.
The SI and SD cross sections were scaled by a factor of $\Omega h^2/0.12$.
It was found that in the $h_1$ scenario, the DM can be either singlino- or higgsino-dominated, whereas in the $h_2$ scenario, it is higgsino-dominated.

Four benchmark points, listed in Tables \ref{Tab2} and \ref{Tab3},  were selected based on their dominant annihilation processes.
These points satisfy the latest direct detection constraints from LZ-2024 \cite{LZ:2024zvo}. Imposing these constraints results in generally small values of the parameter $\lambda$. Within the $h_1$ scenario, points P1 and P2 correspond to singlino-dominated DM, for which the mass satisfies $m_{\tilde{\chi}_1^0} \simeq m_N$, consistent with Eq.(\ref{lightest neutralino mass}).
The dominant annihilation channels differ markedly between P1 and P2.
For point P1 with $m_{\tilde{\chi}_1^0} =$ 23.5 GeV, the dominant DM annihilation channels are $\tilde{\chi}_1^0 \tilde{\chi}_1^0\to b \bar{b}/ \tau^+\tau^-$.
In contrast, for point P2, $m_{\tilde{\chi}_1^0} =$ 146.5 GeV, the condition $2m_{\tilde{\chi}_1^0}>m_{h_s} +m_{a_s}$ is satisfied. This opens the kinematically allowed channel $\tilde{\chi}_1^0 \tilde{\chi}_1^{0} \to h_s a_s$, which becomes the predominant annihilation mode. 
This process proceeds via s-channel exchange of CP-odd Higgs bosons and t-channel exchange of neutralinos.
Given that the contributions from the exchanges of $a_s$ and $\tilde{\chi}_1^0$ are dominant for the small $\lambda$ and sizable $|\kappa|$, the thermally averaged annihilation cross
section $\left\langle \sigma v \right\rangle$ can be simplified as follows \cite{Baum:2017enm}:
\begin{eqnarray} \label{eq:sigvPhiPhi}
\left\langle \sigma v \right\rangle \simeq && \frac{1}{64 \pi m_{\tilde{\chi}_1^0}^2} \left\{ \left[1-\frac{\left(m_{h_s} + m_{a_s}\right)^2}{4 m_{\tilde{\chi}_1^0}^2}\right] \left[1-\frac{\left(m_{h_s} - m_{a_s}\right)^2}{4 m_{\tilde{\chi}_1^0}^2}\right] \right\}^{1/2} | {\cal{A}}_s + {\cal{A}}_t |^2, \nonumber
\end{eqnarray}
where the $s$- and $t$-channel contributions are approximated by
\begin{eqnarray}
{\cal{A}}_s &\simeq & \frac{-2 m_{\tilde{\chi}_1^0} C_{\tilde {\chi}^0_1 \tilde {\chi}^0_1 a_s }  C_{h_s a_s a_s}}{m_{a_s}^2 - 4  m_{\tilde{\chi}_1^0}^2}, \nonumber \\
{\cal{A}}_t &\simeq & - 2 C_{\tilde{\chi}_1^0\tilde{\chi}_1^0 h_s} \, C_{\tilde{\chi}_1^0\tilde{\chi}_1^0 a_s} \left[ 1 + \frac{ 2 m_{a_s}^2}{ 4 m_{\tilde{\chi}_1^0}^2 - \left(m_{h_s}^2 + m_{a_s}^2\right) } \right],
\end{eqnarray}
respectively.
In the small-$\lambda$ case, $C_{\tilde{\chi}_1^0 \tilde{\chi}_1^0 h_s} \simeq C_{\tilde{\chi}_1^0 \tilde{\chi}_1^0 a_s} \simeq -\sqrt{2}\kappa$,
$C_{h_s a_s a_s} \simeq \sqrt{2}\kappa (m_N-A_k)$ \cite{Meng:2024lmi}.
For point P2, $|(m_N - A_\kappa) m_{\tilde{\chi}_1^0}| \ll (4 m_{\tilde{\chi}_1^0}^2 - m_{a_s}^2)$. Therefore, the $t$-channel contributed much more than that from the $s$-channel, which indicates that the t-channel contribution is dominant.
Additionally, the channel $\tilde{\chi}_1^0 \tilde{\chi}_1^{0} \to a_s a_s$ is also open, proceeding through s-channel CP-even Higgs exchange and t-channel neutralino exchange.

Benchmark points P3 and P4 shown in Table \ref{Tab3} represent higgsino-dominated DM in scenarios $h_1$ and $h_2$, respectively. For both points, $N_{13}^2 + N_{14}^2 \simeq 0.994$, which indicates that the DM composition is very close to the pure higgsino limit. 
Table \ref{Tab3} also reveals that the masses of the lightest neutralino, next-to-lightest neutralino, and lightest chargino are nearly degenerate.
The dominant annihilation channels mainly proceed via co‑annihilation processes
$\tilde{\chi}_i^0 \tilde{\chi}_1^{\pm}\to XY$ ($i=1,2$)
and annihilation channels $\tilde{\chi}_1^0 \tilde{\chi}_1^{0} \to W^+W^-/ZZ$,
where X and Y denote the SM quarks or leptons.
The effective annihilation cross-section can be written as ~\cite{Griest:1990kh}
\begin{eqnarray}  
\sigma_{eff} = \sum_{ij} \sigma (\tilde{\chi}_i \tilde{\chi}_j \to X Y) \times \frac{g_i g_j}{g_{eff}^2}(1 + \Delta_i)^{3/2}(1 + \Delta_j)^{3/2}\exp[-x(\Delta_i + \Delta_j)  \label{eff-cross-section}
\end{eqnarray}
where $\Delta_i \equiv (m_{\tilde{\chi}_i} - m_{\tilde{\chi}_1^0})/m_{\tilde{\chi}_1^0}$, $x \equiv m_{\tilde{\chi}_1^0}/T$ (with $T$ being the temperature),
$g_i$ is the number of degrees of freedom for $\tilde{\chi}_i$, and  
\begin{eqnarray}  
g_{eff} \equiv \sum_{i=1}^{N} g_i(1 + \Delta_i)^{3/2}\exp(-x\Delta_i). 
\end{eqnarray}  
Thus, the relic density was primarily determined by the mass splitting among higgsino‑like particles.

Given the complexity of DM physics and its close relation to Higgs phenomenology, a detailed study of DM properties, particularly in the light Higgs scenario, will be the focus of future work.

\section{\label{conclusion}Conclusion}
This study investigated the exotic decay of the SM-like Higgs boson into pairs of the lightest CP-odd or CP-even Higgs bosons within the framework of GNMSSM. 
A comprehensive scan of the parameter space was first conducted using the MultiNest algorithm, subject to experimental constraints from \textsf{HiggsSignals-2.6.2} and \textsf{HiggsBounds-5.10.2}. To assess the implications of recent Higgs precision measurements for light Higgs scenarios (both $h_1=h$ and $h_2=h$ scenarios), 
we separately examined the exclusion limits provided by HiggsSignals and the HiggsBounds implementation in \textsf{HiggsTools}. Within the phenomenologically allowed parameter region, we further conducted a preliminary analysis of DM properties. The main findings were summarized as follows:

\begin{itemize}
\item In both the $h_1$ and $h_2$ scenarios, \textsf{HiggsBounds} imposes the most stringent constraints on the theoretical parameter space. In contrast, \textsf{HiggsSignals} and individual ATLAS searches exhibit comparatively weaker constraining power; moreover, the vast majority of parameter points excluded by these analyses are also disfavored by \textsf{HiggsBounds}. This is because \textsf{HiggsBounds} provides direct constraints from dedicated searches for non-SM Higgs bosons, whereas \textsf{HiggsSignals} derives indirect constraints based on global compatibility with experimental data.

\item Notably, the analysis of the $h_2$ scenario reveals unique phenomenological features. Specifically, \textsf{HiggsSignals} can exclude certain parameter points with relatively low exotic decay rates, especially when $Br(h\to a_1 a_1\to \tau\tau bb) \leq 2.5\%$. This exclusion effect stems from the kinematic accessibility of the $h_2 \to h_1 h_1$ decay process, which has a significantly enhanced branching ratio. Such enhanced decays will modify the properties of the SM-like Higgs boson, thereby increasing the sensitivity of \textsf{HiggsSignals} to indirect constraints in regions where they were previously suppressed due to kinematic factors.

\item Under the combined constraints of \textsf{HiggsTools}, both $h_1$ and $h_2$ scenarios require that the observed $h$ boson is predominantly SM-like ($V_h^{\rm SM} \geq 0.93$) with minimal singlet admixture ($V_h^{\rm S} \leq 0.32$). In the $h_2$ scenario, the lighter $h_s$ state must exhibit high singlet purity ($V_{h_s}^{\rm S} \geq 0.94$) and negligible SM-like components ($V_{h_s}^{\rm SM} \leq 0.32$).

\item In the $h_1$ scenario, DM can be either singlino- or higgsino-dominated, whereas in the $h_2$ scenario, it is higgsino-dominated. For higgsino-dominated DM, mostly annihilation channels of DM are achieved through co-annihilation with chargino; For singlino-dominated DM, mostly annihilation channels of DM may be $\tilde{\chi}_1^0 \tilde{\chi}_1^{0} \to h_s a_s$.
\end{itemize}

Despite constraints from current Higgs data on light Higgs scenarios, phenomenologically viable regions in the parameter space persist. These regions merit further investigation for two principal reasons. First, DM physics exhibits considerable complexity and is deeply intertwined with Higgs phenomenology. Second, the singlet light Higgs spectrum can induce a SFOEWPT in the early universe~\cite{Athron:2019teq,Athron:2023xlk}. This mechanism satisfies the necessary conditions for electroweak baryogenesis (EWBG), providing a plausible explanation for the observed baryon asymmetry of the universe, and simultaneously produces detectable gravitational wave (GW) signals during the phase transition. These aspects will form the core focus of our future work.

\appendix

\section*{Acknowledgement}
We sincerely thank Prof. Junjie Cao for helpful discussions.
We thank LetPub \cite{letpub} for its linguistic assistance during the preparation of this manuscript.
This work was supported by the National Natural Science Foundation of China (Grant No. 12505122) and the Natural Science Foundation of Henan Province (Grant No. 252300423511).


\bibliographystyle{CitationStyle}
\bibliography{main}

@article{ATLAS:2012ae,
  author        = {Aad, Georges and others},
  collaboration = {ATLAS},
  title         = {{Combined search for the Standard Model Higgs boson using up to 4.9 fb$^{-1}$ of $pp$ collision data at $\sqrt{s}=7$ TeV with the ATLAS detector at the LHC}},
  eprint        = {1202.1408},
  archiveprefix = {arXiv},
  primaryclass  = {hep-ex},
  reportnumber  = {CERN-PH-EP-2012-019},
  doi           = {10.1016/j.physletb.2012.02.044},
  journal       = {Phys. Lett. B},
  volume        = {710},
  pages         = {49--66},
  year          = {2012}
}

@article{Barducci:2016pcb,
  author        = {Barducci, D. and Belanger, G. and Bernon, J. and Boudjema, F. and Da Silva, J. and Kraml, S. and Laa, U. and Pukhov, A.},
  title         = {{Collider limits on new physics within micrOMEGAs$\_$4.3}},
  eprint        = {1606.03834},
  archiveprefix = {arXiv},
  primaryclass  = {hep-ph},
  doi           = {10.1016/j.cpc.2017.08.028},
  journal       = {Comput. Phys. Commun.},
  volume        = {222},
  pages         = {327--338},
  year          = {2018}
}

@article{Baum:2017enm,
  author        = {Baum, Sebastian and Carena, Marcela and Shah, Nausheen R. and Wagner, Carlos E.M.},
  title         = {{Higgs portals for thermal Dark Matter. EFT perspectives and the NMSSM}},
  eprint        = {1712.09873},
  archiveprefix = {arXiv},
  primaryclass  = {hep-ph},
  reportnumber  = {NORDITA-2017-130, FERMILAB-PUB-17-611-T, EFI-17-25, WSU-HEP-1715},
  doi           = {10.1007/JHEP04(2018)069},
  journal       = {JHEP},
  volume        = {04},
  pages         = {069},
  year          = {2018}
}

@article{Bechtle:2014ewa,
    author = "Bechtle, Philip and Heinemeyer, Sven and St\r{a}l, Oscar and Stefaniak, Tim and Weiglein, Georg",
    title = "{Probing the Standard Model with Higgs signal rates from the Tevatron, the LHC and a future ILC}",
    eprint = "1403.1582",
    archivePrefix = "arXiv",
    primaryClass = "hep-ph",
    reportNumber = "DESY-14-026, BONN-TH-2014-05",
    doi = "10.1007/JHEP11(2014)039",
    journal = "JHEP",
    volume = "11",
    pages = "039",
    year = "2014"
}

@article{Belanger:2001fz,
  author        = {Belanger, G. and Boudjema, F. and Pukhov, A. and Semenov, A.},
  title         = {{MicrOMEGAs: A Program for calculating the relic density in the MSSM}},
  eprint        = {hep-ph/0112278},
  archiveprefix = {arXiv},
  reportnumber  = {LAPTH-881-01},
  doi           = {10.1016/S0010-4655(02)00596-9},
  journal       = {Comput. Phys. Commun.},
  volume        = {149},
  pages         = {103--120},
  year          = {2002}
}

@article{Belanger:2005kh,
  author        = {Belanger, G. and Boudjema, F. and Hugonie, C. and Pukhov, A. and Semenov, A.},
  title         = {{Relic density of dark matter in the NMSSM}},
  eprint        = {hep-ph/0505142},
  archiveprefix = {arXiv},
  doi           = {10.1088/1475-7516/2005/09/001},
  journal       = {JCAP},
  volume        = {09},
  pages         = {001},
  year          = {2005}
}

@article{Belanger:2006is,
  author        = {Belanger, G. and Boudjema, F. and Pukhov, A. and Semenov, A.},
  title         = {{MicrOMEGAs 2.0: A Program to calculate the relic density of dark matter in a generic model}},
  eprint        = {hep-ph/0607059},
  archiveprefix = {arXiv},
  reportnumber  = {LAPTH-1152-06},
  doi           = {10.1016/j.cpc.2006.11.008},
  journal       = {Comput. Phys. Commun.},
  volume        = {176},
  pages         = {367--382},
  year          = {2007}
}

@article{Belanger:2010pz,
  author        = {Belanger, G. and Boudjema, F. and Pukhov, A. and Semenov, A.},
  title         = {{micrOMEGAs: A Tool for dark matter studies}},
  eprint        = {1005.4133},
  archiveprefix = {arXiv},
  primaryclass  = {hep-ph},
  doi           = {10.1393/ncc/i2010-10591-3},
  journal       = {Nuovo Cim. C},
  volume        = {033N2},
  pages         = {111--116},
  year          = {2010}
}

@article{Belanger:2013oya,
  author        = {Belanger, G. and Boudjema, F. and Pukhov, A. and Semenov, A.},
  title         = {{micrOMEGAs$\_$3: A program for calculating dark matter observables}},
  eprint        = {1305.0237},
  archiveprefix = {arXiv},
  primaryclass  = {hep-ph},
  reportnumber  = {LAPTH-023-13},
  doi           = {10.1016/j.cpc.2013.10.016},
  journal       = {Comput. Phys. Commun.},
  volume        = {185},
  pages         = {960--985},
  year          = {2014}
}

@inproceedings{Belanger:2014hqa,
  author        = {Belanger, G. and Boudjema, F. and Pukhov, A.},
  title         = {{micrOMEGAs : a code for the calculation of Dark Matter
                   properties in generic models of particle interaction}},
  booktitle     = {{The Dark Secrets of the Terascale: Proceedings, TASI
                   2011, Boulder, Colorado, USA, Jun 6 - Jul 11, 2011}},
  year          = {2013},
  pages         = {739-790},
  doi           = {10.1142/9789814390163_0012},
  eprint        = {1402.0787},
  archiveprefix = {arXiv},
  primaryclass  = {hep-ph},
  slaccitation  = {%%CITATION = ARXIV:1402.0787;%%}
}

@article{Belanger:2018mqt,
  author        = {Belanger, Genevieve and Boudjema, Fawzi and Goudelis,
                   Andreas and Pukhov, Alexander and Zaldivar, Bryan},
  title         = {{micrOMEGAs5.0 : Freeze-in}},
  journal       = {Comput. Phys. Commun.},
  volume        = {231},
  year          = {2018},
  pages         = {173-186},
  doi           = {10.1016/j.cpc.2018.04.027},
  eprint        = {1801.03509},
  archiveprefix = {arXiv},
  primaryclass  = {hep-ph},
  slaccitation  = {%%CITATION = ARXIV:1801.03509;%%}
}

@article{Cao:2012fz,
  author        = {Cao, Jun-Jie and Heng, Zhao-Xia and Yang, Jin Min and
                   Zhang, Yan-Ming and Zhu, Jing-Ya},
  title         = {{A SM-like Higgs near 125 GeV in low energy SUSY: a
                   comparative study for MSSM and NMSSM}},
  journal       = {JHEP},
  volume        = {03},
  year          = {2012},
  pages         = {086},
  doi           = {10.1007/JHEP03(2012)086},
  eprint        = {1202.5821},
  archiveprefix = {arXiv},
  primaryclass  = {hep-ph},
  slaccitation  = {%%CITATION = ARXIV:1202.5821;%%}
}

@article{Cao:2016nix,
  author        = {Cao, Junjie and He, Yangle and Shang, Liangliang and Su, Wei and Zhang, Yang},
  title         = {{Natural NMSSM after LHC Run I and the Higgsino dominated dark matter scenario}},
  eprint        = {1606.04416},
  archiveprefix = {arXiv},
  primaryclass  = {hep-ph},
  doi           = {10.1007/JHEP08(2016)037},
  journal       = {JHEP},
  volume        = {08},
  pages         = {037},
  year          = {2016}
}

@article{Cao:2018rix,
  author        = {Cao, Junjie and He, Yangle and Shang, Liangliang and
                   Zhang, Yang and Zhu, Pengxuan},
  title         = {{Current status of a natural NMSSM in light of LHC
                   13 TeV data and XENON-1T results}},
  journal       = {Phys. Rev.},
  volume        = {D99},
  year          = {2019},
  number        = {7},
  pages         = {075020},
  doi           = {10.1103/PhysRevD.99.075020},
  eprint        = {1810.09143},
  archiveprefix = {arXiv},
  primaryclass  = {hep-ph},
  reportnumber  = {CoEPP-MN-18-8},
  slaccitation  = {%%CITATION = ARXIV:1810.09143;%%}
}

@article{Cao:2021ljw,
  author        = {Cao, Junjie and Li, Demin and Lian, Jingwei and Yue, Yuanfang and Zhou, Haijing},
  title         = {{Singlino-dominated dark matter in general NMSSM}},
  eprint        = {2102.05317},
  archiveprefix = {arXiv},
  primaryclass  = {hep-ph},
  month         = {2},
  year          = {2021}
}

@article{Chatrchyan:2012tx,
  author        = {Chatrchyan, Serguei and others},
  collaboration = {CMS},
  title         = {{Combined results of searches for the standard model Higgs boson in $pp$ collisions at $\sqrt{s}=7$ TeV}},
  eprint        = {1202.1488},
  archiveprefix = {arXiv},
  primaryclass  = {hep-ex},
  reportnumber  = {CMS-HIG-11-032, CERN-PH-EP-2012-023},
  doi           = {10.1016/j.physletb.2012.02.064},
  journal       = {Phys. Lett. B},
  volume        = {710},
  pages         = {26--48},
  year          = {2012}
}

@article{Ellwanger:2009dp,
    author = "Ellwanger, Ulrich and Hugonie, Cyril and Teixeira, Ana M.",
    title = "{The Next-to-Minimal Supersymmetric Standard Model}",
    eprint = "0910.1785",
    archivePrefix = "arXiv",
    primaryClass = "hep-ph",
    reportNumber = "LPT-ORSAY-09-76, CFTP-09-032, LPTA-09-066",
    doi = "10.1016/j.physrep.2010.07.001",
    journal = "Phys. Rept.",
    volume = "496",
    pages = "1--77",
    year = "2010"
}

@article{Feroz:2008xx,
  author        = {Feroz, F. and Hobson, M. P. and Bridges, M.},
  title         = {{MultiNest: an efficient and robust Bayesian inference tool for cosmology and particle physics}},
  eprint        = {0809.3437},
  archiveprefix = {arXiv},
  primaryclass  = {astro-ph},
  doi           = {10.1111/j.1365-2966.2009.14548.x},
  journal       = {Mon. Not. Roy. Astron. Soc.},
  volume        = {398},
  pages         = {1601--1614},
  year          = {2009}
}

@article{Gunion:1984yn,
  author       = {Gunion, J. F. and Haber, Howard E.},
  title        = {{Higgs Bosons in Supersymmetric Models. 1.}},
  reportnumber = {SLAC-PUB-3404},
  doi          = {10.1016/0550-3213(86)90340-8},
  journal      = {Nucl. Phys. B},
  volume       = {272},
  pages        = {1},
  year         = {1986},
  note         = {[Erratum: Nucl.Phys.B 402, 567--569 (1993)]}
}

@article{Haber:1984rc,
  author       = {Haber, Howard E. and Kane, Gordon L.},
  title        = {{The Search for Supersymmetry: Probing Physics Beyond the Standard Model}},
  reportnumber = {UM-HE-TH-83-17, SCIPP-85-47},
  doi          = {10.1016/0370-1573(85)90051-1},
  journal      = {Phys. Rept.},
  volume       = {117},
  pages        = {75--263},
  year         = {1985}
}

@article{Martin:1997ns,
  author        = {Martin, Stephen P.},
  editor        = {Kane, Gordon L.},
  title         = {{A Supersymmetry primer}},
  eprint        = {hep-ph/9709356},
  archiveprefix = {arXiv},
  reportnumber  = {FERMILAB-PUB-97-425-T},
  doi           = {10.1142/9789812839657_0001},
  journal       = {Adv. Ser. Direct. High Energy Phys.},
  volume        = {18},
  pages         = {1--98},
  year          = {1998}
}

@article{Meng:2024lmi,
  author        = {Meng, Lei and Cao, Junjie and Li, Fei and Yang, Shenshen},
  title         = {{Dark Matter physics in general NMSSM}},
  eprint        = {2405.07036},
  archiveprefix = {arXiv},
  primaryclass  = {hep-ph},
  doi           = {10.1007/JHEP08(2024)212},
  journal       = {JHEP},
  volume        = {08},
  pages         = {212},
  year          = {2024}
}

@article{Miller:2003ay,
  author        = {Miller, D. J. and Nevzorov, R. and Zerwas, P. M.},
  title         = {{The Higgs sector of the next-to-minimal supersymmetric standard model}},
  eprint        = {hep-ph/0304049},
  archiveprefix = {arXiv},
  reportnumber  = {CERN-TH-2003-077, DESY-03-066, ITEP-5-03},
  doi           = {10.1016/j.nuclphysb.2003.12.021},
  journal       = {Nucl. Phys. B},
  volume        = {681},
  pages         = {3--30},
  year          = {2004}
}

@article{Nilles:1983ge,
  author       = {Nilles, Hans Peter},
  title        = {{Supersymmetry, Supergravity and Particle Physics}},
  reportnumber = {UGVA-DPT-1983-12-412},
  doi          = {10.1016/0370-1573(84)90008-5},
  journal      = {Phys. Rept.},
  volume       = {110},
  pages        = {1--162},
  year         = {1984}
}

@article{Panagiotakopoulos:1998yw,
  author        = {Panagiotakopoulos, C. and Tamvakis, K.},
  title         = {{Stabilized NMSSM without domain walls}},
  eprint        = {hep-ph/9809475},
  archiveprefix = {arXiv},
  doi           = {10.1016/S0370-2693(98)01493-2},
  journal       = {Phys. Lett. B},
  volume        = {446},
  pages         = {224--227},
  year          = {1999}
}

@article{Planck:2018vyg,
  author        = {Aghanim, N. and others},
  collaboration = {Planck},
  title         = {{Planck 2018 results. VI. Cosmological parameters}},
  eprint        = {1807.06209},
  archiveprefix = {arXiv},
  primaryclass  = {astro-ph.CO},
  doi           = {10.1051/0004-6361/201833910},
  journal       = {Astron. Astrophys.},
  volume        = {641},
  pages         = {A6},
  year          = {2020},
  note          = {[Erratum: Astron.Astrophys. 652, C4 (2021)]}
}

@article{Porod_2012,
  title     = {SPheno 3.1: extensions including flavour, CP-phases and models beyond the MSSM},
  volume    = {183},
  issn      = {0010-4655},
  url       = {http://dx.doi.org/10.1016/j.cpc.2012.05.021},
  doi       = {10.1016/j.cpc.2012.05.021},
  number    = {11},
  journal   = {Computer Physics Communications},
  publisher = {Elsevier BV},
  author    = {Porod, W. and Staub, F.},
  year      = {2012},
  month     = {Nov},
  pages     = {2458–2469}
}

@article{Porod:2003um,
  author        = {Porod, Werner},
  title         = {{SPheno, a program for calculating supersymmetric spectra, SUSY particle decays and SUSY particle production at e+ e- colliders}},
  eprint        = {hep-ph/0301101},
  archiveprefix = {arXiv},
  reportnumber  = {ZU-TH-01-03},
  doi           = {10.1016/S0010-4655(03)00222-4},
  journal       = {Comput. Phys. Commun.},
  volume        = {153},
  pages         = {275--315},
  year          = {2003}
}

@article{Pospelov:2007mp,
  author        = {Pospelov, Maxim and Ritz, Adam and Voloshin, Mikhail B.},
  title         = {{Secluded WIMP Dark Matter}},
  eprint        = {0711.4866},
  archiveprefix = {arXiv},
  primaryclass  = {hep-ph},
  doi           = {10.1016/j.physletb.2008.02.052},
  journal       = {Phys. Lett. B},
  volume        = {662},
  pages         = {53--61},
  year          = {2008}
}

@article{Staub:2008uz,
  author        = {Staub, F.},
  title         = {{SARAH}},
  eprint        = {0806.0538},
  archiveprefix = {arXiv},
  primaryclass  = {hep-ph},
  month         = {6},
  year          = {2008}
}

@article{Staub:2012pb,
  author        = {Staub, Florian},
  title         = {{SARAH 3.2: Dirac Gauginos, UFO output, and more}},
  eprint        = {1207.0906},
  archiveprefix = {arXiv},
  primaryclass  = {hep-ph},
  reportnumber  = {BONN-TH-2012-17},
  doi           = {10.1016/j.cpc.2013.02.019},
  journal       = {Comput. Phys. Commun.},
  volume        = {184},
  pages         = {1792--1809},
  year          = {2013}
}

@article{Staub:2013tta,
  author        = {Staub, Florian},
  title         = {{SARAH 4 : A tool for (not only SUSY) model builders}},
  eprint        = {1309.7223},
  archiveprefix = {arXiv},
  primaryclass  = {hep-ph},
  reportnumber  = {BONN-TH-2013-17},
  doi           = {10.1016/j.cpc.2014.02.018},
  journal       = {Comput. Phys. Commun.},
  volume        = {185},
  pages         = {1773--1790},
  year          = {2014}
}

@article{Staub:2015kfa,
  author        = {Staub, Florian},
  title         = {{Exploring new models in all detail with SARAH}},
  eprint        = {1503.04200},
  archiveprefix = {arXiv},
  primaryclass  = {hep-ph},
  reportnumber  = {CERN-PH-TH-2015-051},
  doi           = {10.1155/2015/840780},
  journal       = {Adv. High Energy Phys.},
  volume        = {2015},
  pages         = {840780},
  year          = {2015}
}

@article{Zhou:2021pit,
  author        = {Zhou, Haijing and Cao, Junjie and Lian, Jingwei and Zhang, Di},
  title         = {{Singlino-dominated dark matter in $Z_3$-NMSSM}},
  eprint        = {2102.05309},
  archiveprefix = {arXiv},
  primaryclass  = {hep-ph},
  month         = {2},
  year          = {2021}
}

@article{LZ:2024zvo,
  author        = {Aalbers, J. and others},
  collaboration = {LZ},
  title         = {{Dark Matter Search Results from 4.2 Tonne-Years of Exposure of the LUX-ZEPLIN (LZ) Experiment}},
  eprint        = {2410.17036},
  archiveprefix = {arXiv},
  primaryclass  = {hep-ex},
  doi           = {10.1103/4dyc-z8zf},
  journal       = {Phys. Rev. Lett.},
  volume        = {135},
  number        = {1},
  pages         = {011802},
  reportnumber  = {FERMILAB-PUB-24-0796-V},
  month         = {10},
  year          = {2025}
}

@article{LZ:2022lsv,
    author = "Aalbers, J. and others",
    collaboration = "LZ",
    title = "{First Dark Matter Search Results from the LUX-ZEPLIN (LZ) Experiment}",
    eprint = "2207.03764",
    archivePrefix = "arXiv",
    primaryClass = "hep-ex",
    doi = "10.1103/PhysRevLett.131.041002",
    journal = "Phys. Rev. Lett.",
    volume = "131",
    number = "4",
    pages = "041002",
    year = "2023"
}

@article{Maniatis:2009re,
    author = "Maniatis, M.",
    title = "{The Next-to-Minimal Supersymmetric extension of the Standard Model reviewed}",
    eprint = "0906.0777",
    archivePrefix = "arXiv",
    primaryClass = "hep-ph",
    reportNumber = "HD-THEP-09-9",
    doi = "10.1142/S0217751X10049827",
    journal = "Int. J. Mod. Phys. A",
    volume = "25",
    pages = "3505--3602",
    year = "2010"
}

@article{Cao:2024axg,
    author = "Cao, Junjie and Jia, Xinglong and Lian, Jingwei",
    title = "{Unified interpretation of the muon g-2 anomaly, the 95~GeV diphoton, and $b\bar b$ excesses in the general next-to-minimal supersymmetric standard model}",
    eprint = "2402.15847",
    archivePrefix = "arXiv",
    primaryClass = "hep-ph",
    doi = "10.1103/PhysRevD.110.115039",
    journal = "Phys. Rev. D",
    volume = "110",
    number = "11",
    pages = "115039",
    year = "2024"
}

@article{Cao:2022ovk,
    author = "Cao, Junjie and Jia, Xinglong and Meng, Lei and Yue, Yuanfang and Zhang, Di",
    title = "{Status of the singlino-dominated dark matter in general Next-to-Minimal Supersymmetric Standard Model}",
    eprint = "2210.08769",
    archivePrefix = "arXiv",
    primaryClass = "hep-ph",
    doi = "10.1007/JHEP03(2023)198",
    journal = "JHEP",
    volume = "03",
    pages = "198",
    year = "2023"
}

@article{Abel:1996cr,
    author = "Abel, S. A.",
    title = "{Destabilizing divergences in the NMSSM}",
    eprint = "hep-ph/9609323",
    archivePrefix = "arXiv",
    reportNumber = "ULB-TH-96-16",
    doi = "10.1016/S0550-3213(96)00470-1",
    journal = "Nucl. Phys. B",
    volume = "480",
    pages = "55--72",
    year = "1996"
}

@article{Kolda:1998rm,
    author = "Kolda, Christopher F. and Pokorski, Stefan and Polonsky, Nir",
    title = "{Stabilized singlets in supergravity as a source of the mu - parameter}",
    eprint = "hep-ph/9803310",
    archivePrefix = "arXiv",
    reportNumber = "IASSNS-HEP-97-137, CERN-TH-98-75, RU-97-97",
    doi = "10.1103/PhysRevLett.80.5263",
    journal = "Phys. Rev. Lett.",
    volume = "80",
    pages = "5263--5266",
    year = "1998"
}

@article{Ross:2011xv,
    author = "Ross, Graham G. and Schmidt-Hoberg, Kai",
    title = "{The Fine-Tuning of the Generalised NMSSM}",
    eprint = "1108.1284",
    archivePrefix = "arXiv",
    primaryClass = "hep-ph",
    reportNumber = "OUTP-11-48P",
    doi = "10.1016/j.nuclphysb.2012.05.007",
    journal = "Nucl. Phys. B",
    volume = "862",
    pages = "710--719",
    year = "2012"
}

@article{Lee:2010gv,
    author = "Lee, Hyun Min and Raby, Stuart and Ratz, Michael and Ross, Graham G. and Schieren, Roland and Schmidt-Hoberg, Kai and Vaudrevange, Patrick K. S.",
    title = "{A unique $\mathbb{Z}_4^R$ symmetry for the MSSM}",
    eprint = "1009.0905",
    archivePrefix = "arXiv",
    primaryClass = "hep-ph",
    reportNumber = "TUM-HEP-770-10, LMU-ASC-64-10, OHSTPY-HEP-T-10-003, CERN-PH-TH-2010-193, OUTP-10-24P",
    doi = "10.1016/j.physletb.2010.10.038",
    journal = "Phys. Lett. B",
    volume = "694",
    pages = "491--495",
    year = "2011"
}

@article{Lee:2011dya,
    author = "Lee, Hyun Min and Raby, Stuart and Ratz, Michael and Ross, Graham G. and Schieren, Roland and Schmidt-Hoberg, Kai and Vaudrevange, Patrick K. S.",
    title = "{Discrete R symmetries for the MSSM and its singlet extensions}",
    eprint = "1102.3595",
    archivePrefix = "arXiv",
    primaryClass = "hep-ph",
    reportNumber = "TUM-HEP-793-11, LMU-ASC-06-11, OHSTPY-HEP-T-11-001, CERN-PH-TH-2011-022, OUTP-11-33P",
    doi = "10.1016/j.nuclphysb.2011.04.009",
    journal = "Nucl. Phys. B",
    volume = "850",
    pages = "1--30",
    year = "2011"
}

@article{Ross:2012nr,
    author = "Ross, Graham G. and Schmidt-Hoberg, Kai and Staub, Florian",
    title = "{The Generalised NMSSM at One Loop: Fine Tuning and Phenomenology}",
    eprint = "1205.1509",
    archivePrefix = "arXiv",
    primaryClass = "hep-ph",
    reportNumber = "OUTP-12-06P, BONN-TH-2012-04",
    doi = "10.1007/JHEP08(2012)074",
    journal = "JHEP",
    volume = "08",
    pages = "074",
    year = "2012"
}

@article{Ellwanger:1983mg,
    author = "Ellwanger, U.",
    title = "{NONRENORMALIZABLE INTERACTIONS FROM SUPERGRAVITY, QUANTUM CORRECTIONS AND EFFECTIVE LOW-ENERGY THEORIES}",
    doi = "10.1016/0370-2693(83)90557-9",
    journal = "Phys. Lett. B",
    volume = "133",
    pages = "187--191",
    year = "1983"
}

@article{ATLAS:2020zms,
    author = "Aad, Georges and others",
    collaboration = "ATLAS",
    title = "{Search for heavy Higgs bosons decaying into two tau leptons with the ATLAS detector using $pp$ collisions at $\sqrt{s}=13$ TeV}",
    eprint = "2002.12223",
    archivePrefix = "arXiv",
    primaryClass = "hep-ex",
    reportNumber = "CERN-EP-2020-014",
    doi = "10.1103/PhysRevLett.125.051801",
    journal = "Phys. Rev. Lett.",
    volume = "125",
    number = "5",
    pages = "051801",
    year = "2020"
}

@article{ATLAS:2021upq,
    author = "Aad, Georges and others",
    collaboration = "ATLAS",
    title = "{Search for charged Higgs bosons decaying into a top quark and a bottom quark at $ \sqrt{\mathrm{s}} $ = 13 TeV with the ATLAS detector}",
    eprint = "2102.10076",
    archivePrefix = "arXiv",
    primaryClass = "hep-ex",
    reportNumber = "CERN-EP-2021-004",
    doi = "10.1007/JHEP06(2021)145",
    journal = "JHEP",
    volume = "06",
    pages = "145",
    year = "2021"
}

@article{Griest:1990kh,
    author = "Griest, Kim and Seckel, David",
    title = "{Three exceptions in the calculation of relic abundances}",
    reportNumber = "CFPA-TH-90-001A, BA-90-79",
    doi = "10.1103/PhysRevD.43.3191",
    journal = "Phys. Rev. D",
    volume = "43",
    pages = "3191--3203",
    year = "1991"
}

@article{Bahl:2022igd,
    author = {Bahl, Henning and Biek\"otter, Thomas and Heinemeyer, Sven and Li, Cheng and Paasch, Steven and Weiglein, Georg and Wittbrodt, Jonas},
    title = "{HiggsTools: BSM scalar phenomenology with new versions of HiggsBounds and HiggsSignals}",
    eprint = "2210.09332",
    archivePrefix = "arXiv",
    primaryClass = "hep-ph",
    doi = "10.1016/j.cpc.2023.108803",
    journal = "Comput. Phys. Commun.",
    volume = "291",
    pages = "108803",
    year = "2023"
}

@article{pdg2020,
    author = "Zyla, P.A. and others",
    collaboration = "Particle Data Group",
    title = "{Review of Particle Physics}",
    doi = "10.1093/ptep/ptaa104",
    journal = "PTEP",
    volume = "2020",
    number = "8",
    pages = "083C01",
    year = "2020"
}

@article{HB2008jh,
    author = "Bechtle, Philip and Brein, Oliver and Heinemeyer, Sven and Weiglein, Georg and Williams, Karina E.",
    title = "{HiggsBounds: Confronting Arbitrary Higgs Sectors with Exclusion Bounds from LEP and the Tevatron}",
    eprint = "0811.4169",
    archivePrefix = "arXiv",
    primaryClass = "hep-ph",
    reportNumber = "DCPT-08-172, IPPP-08-86, BONN-TH-2008-17",
    doi = "10.1016/j.cpc.2009.09.003",
    journal = "Comput. Phys. Commun.",
    volume = "181",
    pages = "138--167",
    year = "2010"
}

@article{HB2011sb,
    author = "Bechtle, Philip and Brein, Oliver and Heinemeyer, Sven and Weiglein, Georg and Williams, Karina E.",
    title = "{HiggsBounds 2.0.0: Confronting Neutral and Charged Higgs Sector Predictions with Exclusion Bounds from LEP and the Tevatron}",
    eprint = "1102.1898",
    archivePrefix = "arXiv",
    primaryClass = "hep-ph",
    reportNumber = "FR-PHENO-2011-002, BONN-TH-2011-02, DESY-11-016",
    doi = "10.1016/j.cpc.2011.07.015",
    journal = "Comput. Phys. Commun.",
    volume = "182",
    pages = "2605--2631",
    year = "2011"
}

@article{HBHS2012lvg,
    author = "Bechtle, Philip and Brein, Oliver and Heinemeyer, Sven and Stal, Oscar and Stefaniak, Tim and Weiglein, Georg and Williams, Karina",
    editor = "Enberg, Rikard and Ferrari, Arnaud",
    title = "{Recent Developments in HiggsBounds and a Preview of HiggsSignals}",
    eprint = "1301.2345",
    archivePrefix = "arXiv",
    primaryClass = "hep-ph",
    reportNumber = "BONN-TH-2013-01, DESY-13-004",
    doi = "10.22323/1.156.0024",
    journal = "PoS",
    volume = "CHARGED2012",
    pages = "024",
    year = "2012"
}

@article{HB2013wla,
    author = "Bechtle, Philip and Brein, Oliver and Heinemeyer, Sven and Stal, Oscar and Stefaniak, Tim and Weiglein, Georg and Williams, Karina E.",
    title = "{$\mathsf{HiggsBounds}-4$: Improved Tests of Extended Higgs Sectors against Exclusion Bounds from LEP, the Tevatron and the LHC}",
    eprint = "1311.0055",
    archivePrefix = "arXiv",
    primaryClass = "hep-ph",
    reportNumber = "BONN-TH-2013-21, DESY-13-110",
    doi = "10.1140/epjc/s10052-013-2693-2",
    journal = "Eur. Phys. J. C",
    volume = "74",
    number = "3",
    pages = "2693",
    year = "2014"
}

@article{HB2020pkv,
    author = "Bechtle, Philip and Dercks, Daniel and Heinemeyer, Sven and Klingl, Tobias and Stefaniak, Tim and Weiglein, Georg and Wittbrodt, Jonas",
    title = "{HiggsBounds-5: Testing Higgs Sectors in the LHC 13 TeV Era}",
    eprint = "2006.06007",
    archivePrefix = "arXiv",
    primaryClass = "hep-ph",
    reportNumber = "BONN-TH-2020-03, DESY 20-093, DESY-20-093, IFT-UAM/CSIC-20-072, LU 20-27",
    doi = "10.1140/epjc/s10052-020-08557-9",
    journal = "Eur. Phys. J. C",
    volume = "80",
    number = "12",
    pages = "1211",
    year = "2020"
}

@article{HS2013xfa,
    author = "Bechtle, Philip and Heinemeyer, Sven and St\r{a}l, Oscar and Stefaniak, Tim and Weiglein, Georg",
    title = "{$HiggsSignals$: Confronting arbitrary Higgs sectors with measurements at the Tevatron and the LHC}",
    eprint = "1305.1933",
    archivePrefix = "arXiv",
    primaryClass = "hep-ph",
    reportNumber = "BONN-TH-2013-07, DESY-13-078",
    doi = "10.1140/epjc/s10052-013-2711-4",
    journal = "Eur. Phys. J. C",
    volume = "74",
    number = "2",
    pages = "2711",
    year = "2014"
}

@article{HSConstraining2013hwa,
    author = "St\r{a}l, Oscar and Stefaniak, Tim",
    title = "{Constraining extended Higgs sectors with HiggsSignals}",
    eprint = "1310.4039",
    archivePrefix = "arXiv",
    primaryClass = "hep-ph",
    reportNumber = "BONN-TH-2013-20",
    doi = "10.22323/1.180.0314",
    journal = "PoS",
    volume = "EPS-HEP2013",
    pages = "314",
    year = "2013"
}

@article{HS2020uwn,
    author = "Bechtle, Philip and Heinemeyer, Sven and Klingl, Tobias and Stefaniak, Tim and Weiglein, Georg and Wittbrodt, Jonas",
    title = "{HiggsSignals-2: Probing new physics with precision Higgs measurements in the LHC 13 TeV era}",
    eprint = "2012.09197",
    archivePrefix = "arXiv",
    primaryClass = "hep-ph",
    reportNumber = "BONN-TH-2020-09, DESY-20-228, DESY 20-228, IFT-UAM/CSIC-20-081, LU TP 20-53",
    doi = "10.1140/epjc/s10052-021-08942-y",
    journal = "Eur. Phys. J. C",
    volume = "81",
    number = "2",
    pages = "145",
    year = "2021"
}

@article{CMS2024,
    author = "Hayrapetyan, Aram and others",
    collaboration = "CMS",
    title = "{Search for exotic decays of the Higgs boson to a pair of pseudoscalars in the $\mu\mu$bb and $\tau\tau$bb final states}",
    eprint = "2402.13358",
    archivePrefix = "arXiv",
    primaryClass = "hep-ex",
    reportNumber = "CMS-HIG-22-007, CERN-EP-2023-284",
    doi = "10.1140/epjc/s10052-024-12727-4",
    journal = "Eur. Phys. J. C",
    volume = "84",
    number = "5",
    pages = "493",
    year = "2024"
}

@article{ATLAS2024,
    author = "Aad, Georges and others",
    collaboration = "ATLAS",
    title = "{Search for decays of the Higgs boson into a pair of pseudoscalar particles decaying into bb\textasciimacron{}\ensuremath{\tau}+\ensuremath{\tau}- using pp collisions at s=13\,\,TeV with the ATLAS detector}",
    eprint = "2407.01335",
    archivePrefix = "arXiv",
    primaryClass = "hep-ex",
    reportNumber = "CERN-EP-2024-164",
    doi = "10.1103/PhysRevD.110.052013",
    journal = "Phys. Rev. D",
    volume = "110",
    number = "5",
    pages = "052013",
    year = "2024"
}

@article{95GeV2023,
    author = "Cao, Junjie and Jia, Xinglong and Lian, Jingwei and Meng, Lei",
    title = "{95~GeV diphoton and $b \bar b$ excesses in the general next-to-minimal supersymmetric standard model}",
    eprint = "2310.08436",
    archivePrefix = "arXiv",
    primaryClass = "hep-ph",
    doi = "10.1103/PhysRevD.109.075001",
    journal = "Phys. Rev. D",
    volume = "109",
    number = "7",
    pages = "075001",
    year = "2024"
}

@article{Cao:2016cnv,
    author = "Cao, Junjie and He, Yangle and Shang, Liangliang and Su, Wei and Wu, Peiwen and Zhang, Yang",
    title = "{Strong constraints of LUX-2016 results on the natural NMSSM}",
    eprint = "1609.00204",
    archivePrefix = "arXiv",
    primaryClass = "hep-ph",
    doi = "10.1007/JHEP10(2016)136",
    journal = "JHEP",
    volume = "10",
    pages = "136",
    year = "2016"
}

@article{Heng:2018kyd,
    author = "Heng, Zhaoxia and Gong, Xue and Zhou, Haijing",
    title = "{Pair production of Higgs boson in NMSSM at the LHC with the next-to-lightest CP-even Higgs boson being SM-like}",
    eprint = "1805.01598",
    archivePrefix = "arXiv",
    primaryClass = "hep-ph",
    doi = "10.1088/1674-1137/42/7/073103",
    journal = "Chin. Phys. C",
    volume = "42",
    number = "7",
    pages = "073103",
    year = "2018"
}

@article{Cao:2016uwt,
    author = "Cao, Junjie and Guo, Xiaofei and He, Yangle and Wu, Peiwen and Zhang, Yang",
    title = "{Diphoton signal of the light Higgs boson in natural NMSSM}",
    eprint = "1612.08522",
    archivePrefix = "arXiv",
    primaryClass = "hep-ph",
    doi = "10.1103/PhysRevD.95.116001",
    journal = "Phys. Rev. D",
    volume = "95",
    number = "11",
    pages = "116001",
    year = "2017"
}

@article{Cao:2014kya,
    author = "Cao, Junjie and Li, Dongwei and Shang, Liangliang and Wu, Peiwen and Zhang, Yang",
    title = "{Exploring the Higgs Sector of a Most Natural NMSSM and its Prediction on Higgs Pair Production at the LHC}",
    eprint = "1409.8431",
    archivePrefix = "arXiv",
    primaryClass = "hep-ph",
    doi = "10.1007/JHEP12(2014)026",
    journal = "JHEP",
    volume = "12",
    pages = "026",
    year = "2014"
}

@article{Ellwanger:2024etv,
    author = "Ellwanger, Ulrich and Muehlleitner, Margarete and Rompotis, Nikolaos and Shah, Nausheen R. and Winterbottom, Daniel",
    title = "{Benchmark Lines and Planes for Higgs-to-Higgs Decays in the NMSSM}",
    eprint = "2403.15046",
    archivePrefix = "arXiv",
    primaryClass = "hep-ph",
    reportNumber = "LHCHWG-2024-002",
    month = "3",
    year = "2024"
}

@article{Ellwanger:2022jtd,
    author = "Ellwanger, Ulrich and Hugonie, Cyril",
    title = "{Benchmark planes for Higgs-to-Higgs decays in the NMSSM}",
    eprint = "2203.05049",
    archivePrefix = "arXiv",
    primaryClass = "hep-ph",
    reportNumber = "LUPM 22-004",
    doi = "10.1140/epjc/s10052-022-10364-3",
    journal = "Eur. Phys. J. C",
    volume = "82",
    number = "5",
    pages = "406",
    year = "2022"
}

@article{Ma:2020mjz,
    author = "Ma, Shiquan and Wang, Kun and Zhu, Jingya",
    title = "{Higgs decay to light (pseudo)scalars in the semi-constrained NMSSM}",
    eprint = "2006.03527",
    archivePrefix = "arXiv",
    primaryClass = "hep-ph",
    reportNumber = "WHU-HEP-PH-TEV008",
    doi = "10.1088/1674-1137/abce4f",
    journal = "Chin. Phys. C",
    volume = "45",
    number = "2",
    pages = "023113",
    year = "2021"
}

@article{Wang:2016lvj,
    author = "Wang, Wenyu and Zhang, Mengchao and Zhao, Jun",
    title = "{Higgs exotic decays in general NMSSM with self-interacting dark matter}",
    eprint = "1604.00123",
    archivePrefix = "arXiv",
    primaryClass = "hep-ph",
    doi = "10.1142/S0217751X18410026",
    journal = "Int. J. Mod. Phys. A",
    volume = "33",
    number = "11",
    pages = "1841002",
    year = "2018"
}

@article{Cao:2013gba,
    author = "Cao, Junjie and Ding, Fangfang and Han, Chengcheng and Yang, Jin Min and Zhu, Jingya",
    title = "{A light Higgs scalar in the NMSSM confronted with the latest LHC Higgs data}",
    eprint = "1309.4939",
    archivePrefix = "arXiv",
    primaryClass = "hep-ph",
    doi = "10.1007/JHEP11(2013)018",
    journal = "JHEP",
    volume = "11",
    pages = "018",
    year = "2013"
}

@article{Curtin:2013fra,
    author = "Curtin, David and others",
    title = "{Exotic decays of the 125 GeV Higgs boson}",
    eprint = "1312.4992",
    archivePrefix = "arXiv",
    primaryClass = "hep-ph",
    reportNumber = "YITP-13-47, PITT-PACC-1314",
    doi = "10.1103/PhysRevD.90.075004",
    journal = "Phys. Rev. D",
    volume = "90",
    number = "7",
    pages = "075004",
    year = "2014"
}

@article{Cheng:2024gfs,
    author = "Cheng, Junyi and Husain, Rabia and Li, Lingfeng and Strassler, Matthew J.",
    title = "{Limits on an Exotic Higgs Decay From a Recast ATLAS Four-Lepton Analysis}",
    eprint = "2412.14452",
    archivePrefix = "arXiv",
    primaryClass = "hep-ph",
    month = "12",
    year = "2024"
}

@article{Zhou:2025xol,
    author = "Zhou, Haijing and Ban, Guangning",
    title = "{Status of $\mathbb{Z}_3$-NMSSM featuring a light bino-dominated LSP and a light singlet-like scalar under the LZ Experiment}",
    eprint = "2502.14664",
    archivePrefix = "arXiv",
    primaryClass = "hep-ph",
    month = "2",
    year = "2025"
}

@article{Huang:2013ima,
    author = "Huang, Jinrui and Liu, Tao and Wang, Lian-Tao and Yu, Felix",
    title = "{Supersymmetric Exotic Decays of the 125 GeV Higgs Boson}",
    eprint = "1309.6633",
    archivePrefix = "arXiv",
    primaryClass = "hep-ph",
    reportNumber = "FERMILAB-PUB-13-186-T, LA-UR-13-24173",
    doi = "10.1103/PhysRevLett.112.221803",
    journal = "Phys. Rev. Lett.",
    volume = "112",
    number = "22",
    pages = "221803",
    year = "2014"
}

@article{Cao:2022chy,
    author = "Cao, Junjie and Lian, Jingwei and Pan, Yusi and Yue, Yuanfang and Zhang, Di",
    title = "{Impact of recent (g \ensuremath{-} 2)$_{\mu}$ measurement on the light CP-even Higgs scenario in general Next-to-Minimal Supersymmetric Standard Model}",
    eprint = "2201.11490",
    archivePrefix = "arXiv",
    primaryClass = "hep-ph",
    doi = "10.1007/JHEP03(2022)203",
    journal = "JHEP",
    volume = "03",
    pages = "203",
    year = "2022"
}

@article{CMS:2024zfv,
    author = "Hayrapetyan, Aram and others",
    collaboration = "CMS",
    title = "{Search for the decay of the Higgs boson to a pair of light pseudoscalar bosons in the final state with four bottom quarks in proton-proton collisions at $ \sqrt{\textrm{s}} $ = 13 TeV}",
    eprint = "2403.10341",
    archivePrefix = "arXiv",
    primaryClass = "hep-ex",
    reportNumber = "CMS-HIG-18-026, CERN-EP-2024-028",
    doi = "10.1007/JHEP06(2024)097",
    journal = "JHEP",
    volume = "06",
    pages = "097",
    year = "2024"
}

@article{CMS:2017dmg,
    author = "Khachatryan, V. and others",
    collaboration = "CMS",
    title = "{Search for light bosons in decays of the 125 GeV Higgs boson in proton-proton collisions at $ \sqrt{s}=8 $ TeV}",
    eprint = "1701.02032",
    archivePrefix = "arXiv",
    primaryClass = "hep-ex",
    reportNumber = "CMS-HIG-16-015, CERN-EP-2016-292",
    doi = "10.1007/JHEP10(2017)076",
    journal = "JHEP",
    volume = "10",
    pages = "076",
    year = "2017"
}

@article{ATLAS:2018emt,
    author = "Aaboud, Morad and others",
    collaboration = "ATLAS",
    title = "{Search for Higgs boson decays into a pair of light bosons in the $bb\mu\mu$ final state in $pp$ collision at $\sqrt{s} = $13 TeV with the ATLAS detector}",
    eprint = "1807.00539",
    archivePrefix = "arXiv",
    primaryClass = "hep-ex",
    reportNumber = "CERN-EP-2018-153",
    doi = "10.1016/j.physletb.2018.10.073",
    journal = "Phys. Lett. B",
    volume = "790",
    pages = "1--21",
    year = "2019"
}

@article{CMS:2018zvv,
    author = "Sirunyan, Albert M and others",
    collaboration = "CMS",
    title = "{Search for an exotic decay of the Higgs boson to a pair of light pseudoscalars in the final state with two b quarks and two $\tau$ leptons in proton-proton collisions at $\sqrt{s}=$ 13 TeV}",
    eprint = "1805.10191",
    archivePrefix = "arXiv",
    primaryClass = "hep-ex",
    reportNumber = "CMS-HIG-17-024, CERN-EP-2018-089",
    doi = "10.1016/j.physletb.2018.08.057",
    journal = "Phys. Lett. B",
    volume = "785",
    pages = "462",
    year = "2018"
}

@article{ATLAS:2024nnm,
    author = "Aad, Georges and others",
    collaboration = "ATLAS",
    title = "{Search for Higgs boson decays into a pair of pseudoscalar particles in the $\gamma\gamma\tau_{\text{had}}\tau_{\text{had}}$ final state using $pp$ collisions at $\sqrt{s}=13$ TeV with the ATLAS detector}",
    eprint = "2412.14046",
    archivePrefix = "arXiv",
    primaryClass = "hep-ex",
    reportNumber = "CERN-EP-2024-323",
    month = "12",
    year = "2024"
}

@article{ATLAS:2015hpr,
    author = "Aad, Georges and others",
    collaboration = "ATLAS",
    title = "{Search for new light gauge bosons in Higgs boson decays to four-lepton final states in $pp$ collisions at $\sqrt{s}=8$ TeV with the ATLAS detector at the LHC}",
    eprint = "1505.07645",
    archivePrefix = "arXiv",
    primaryClass = "hep-ex",
    reportNumber = "CERN-PH-EP-2015-111",
    doi = "10.1103/PhysRevD.92.092001",
    journal = "Phys. Rev. D",
    volume = "92",
    number = "9",
    pages = "092001",
    year = "2015"
}

@article{CMS:2012qms,
    author = "Chatrchyan, Serguei and others",
    collaboration = "CMS",
    title = "{Search for a Non-Standard-Model Higgs Boson Decaying to a Pair of New Light Bosons in Four-Muon Final States}",
    eprint = "1210.7619",
    archivePrefix = "arXiv",
    primaryClass = "hep-ex",
    reportNumber = "CMS-EXO-12-012, CERN-PH-EP-2012-292",
    doi = "10.1016/j.physletb.2013.09.009",
    journal = "Phys. Lett. B",
    volume = "726",
    pages = "564--586",
    year = "2013"
}

@article{CMS:2015nay,
    author = "Khachatryan, V. and others",
    collaboration = "CMS",
    title = "{A search for pair production of new light bosons decaying into muons}",
    eprint = "1506.00424",
    archivePrefix = "arXiv",
    primaryClass = "hep-ex",
    reportNumber = "CMS-HIG-13-010, CERN-PH-EP-2015-116",
    doi = "10.1016/j.physletb.2015.10.067",
    journal = "Phys. Lett. B",
    volume = "752",
    pages = "146--168",
    year = "2016"
}

@article{ATLAS:2018coo,
    author = "Aaboud, Morad and others",
    collaboration = "ATLAS",
    title = "{Search for Higgs boson decays to beyond-the-Standard-Model light bosons in four-lepton events with the ATLAS detector at $\sqrt{s}=13$ TeV}",
    eprint = "1802.03388",
    archivePrefix = "arXiv",
    primaryClass = "hep-ex",
    reportNumber = "CERN-EP-2017-293",
    doi = "10.1007/JHEP06(2018)166",
    journal = "JHEP",
    volume = "06",
    pages = "166",
    year = "2018"
}

@article{CMS:2018jid,
    author = "Sirunyan, Albert M and others",
    collaboration = "CMS",
    title = "{A search for pair production of new light bosons decaying into muons in proton-proton collisions at 13 TeV}",
    eprint = "1812.00380",
    archivePrefix = "arXiv",
    primaryClass = "hep-ex",
    reportNumber = "CMS-HIG-18-003, CERN-EP-2018-288",
    doi = "10.1016/j.physletb.2019.07.013",
    journal = "Phys. Lett. B",
    volume = "796",
    pages = "131--154",
    year = "2019"
}

@article{letpub,
author="https://www.letpub.com.cn/"
}

@article{XENON:2018voc,
    author = "Aprile, E. and others",
    collaboration = "XENON",
    title = "{Dark Matter Search Results from a One Ton-Year Exposure of XENON1T}",
    eprint = "1805.12562",
    archivePrefix = "arXiv",
    primaryClass = "astro-ph.CO",
    doi = "10.1103/PhysRevLett.121.111302",
    journal = "Phys. Rev. Lett.",
    volume = "121",
    number = "11",
    pages = "111302",
    year = "2018"
}

@article{PandaX-II:2020oim,
    author = "Wang, Qiuhong and others",
    collaboration = "PandaX-II",
    title = "{Results of dark matter search using the full PandaX-II exposure}",
    eprint = "2007.15469",
    archivePrefix = "arXiv",
    primaryClass = "astro-ph.CO",
    doi = "10.1088/1674-1137/abb658",
    journal = "Chin. Phys. C",
    volume = "44",
    number = "12",
    pages = "125001",
    year = "2020"
}

@article{PandaX-II:2017hlx,
    author = "Cui, Xiangyi and others",
    collaboration = "PandaX-II",
    title = "{Dark Matter Results From 54-Ton-Day Exposure of PandaX-II Experiment}",
    eprint = "1708.06917",
    archivePrefix = "arXiv",
    primaryClass = "astro-ph.CO",
    doi = "10.1103/PhysRevLett.119.181302",
    journal = "Phys. Rev. Lett.",
    volume = "119",
    number = "18",
    pages = "181302",
    year = "2017"
}

@article{Cao:2022htd,
    author = "Cao, Junjie and Li, Fei and Lian, Jingwei and Pan, Yusi and Zhang, Di",
    title = "{Impact of LHC probes of SUSY and recent measurement of $(g-2)_{\mu}$ on {$Z_{3}$-NMSSM}}",
    eprint = "2204.04710",
    archivePrefix = "arXiv",
    primaryClass = "hep-ph",
    doi = "10.1007/s11433-022-1927-9",
    journal = "Sci. China Phys. Mech. Astron.",
    volume = "65",
    number = "9",
    pages = "291012",
    year = "2022"
}

@article{Athron:2019teq,
    author = "Athron, Peter and Balazs, Csaba and Fowlie, Andrew and Pozzo, Giancarlo and White, Graham and Zhang, Yang",
    title = "{Strong first-order phase transitions in the NMSSM-a comprehensive survey}",
    eprint = "1908.11847",
    archivePrefix = "arXiv",
    primaryClass = "hep-ph",
    reportNumber = "CoEPP-MN-19-03",
    doi = "10.1007/JHEP11(2019)151",
    journal = "JHEP",
    volume = "11",
    pages = "151",
    year = "2019"
}

@article{Athron:2023xlk,
    author = "Athron, Peter and Balazs, Csaba and Fowlie, Andrew and Morris, Lachlan and Wu, Lei",
    title = "{Cosmological phase transitions: From perturbative particle physics to gravitational waves}",
    eprint = "2305.02357",
    archivePrefix = "arXiv",
    primaryClass = "hep-ph",
    doi = "10.1016/j.ppnp.2023.104094",
    journal = "Prog. Part. Nucl. Phys.",
    volume = "135",
    pages = "104094",
    year = "2024"
}

@article{ATLAS:2015yey,
    author = "G.~Aad \textit{et al.} [ATLAS and CMS]",
    title = "{Combined Measurement of the Higgs Boson Mass in $pp$ Collisions at $\sqrt{s}=7$ and 8 TeV with the ATLAS and CMS Experiments}",
    eprint = "1503.07589 ",
    archivePrefix = "arXiv",
    primaryClass = "hep-ex",
    doi = "doi:10.1103/PhysRevLett.114.191803",
    journal = " Phys. Rev. Lett.", 
    volume = "114",
    pages = "191803",
    year = "2015"
}

@article{Liu:2016ahc,
    author = "Liu, Shang and Tang, Yi-Lei and Zhang, Chen and Zhu, Shou-hua",
    title = "{Exotic Higgs Decay $h\rightarrow\phi\phi\rightarrow 4b$ at the LHeC}",
    eprint = "1608.08458",
    archivePrefix = "arXiv",
    primaryClass = "hep-ph",
    doi = "10.1140/epjc/s10052-017-5012-5",
    journal = "Eur. Phys. J. C",
    volume = "77",
    number = "7",
    pages = "457",
    year = "2017"
}

@article{Domingo:2016unq,
    author = "Domingo, F. and Heinemeyer, S. and Kim, J. S. and Rolbiecki, K.",
    title = "{The NMSSM lives: with the 750 GeV diphoton excess}",
    eprint = "1602.07691",
    archivePrefix = "arXiv",
    primaryClass = "hep-ph",
    reportNumber = "IFT-UAM-CSIC-16-019",
    doi = "10.1140/epjc/s10052-016-4080-2",
    journal = "Eur. Phys. J. C",
    volume = "76",
    number = "5",
    pages = "249",
    year = "2016"
}

@article{Curtin:2014pda,
    author = "Curtin, David and Essig, Rouven and Zhong, Yi-Ming",
    title = "{Uncovering light scalars with exotic Higgs decays to $ b\overline{b}{\mu}^{+}{\mu}^{-} $}",
    eprint = "1412.4779",
    archivePrefix = "arXiv",
    primaryClass = "hep-ph",
    reportNumber = "YITP-SB-14-53",
    doi = "10.1007/JHEP06(2015)025",
    journal = "JHEP",
    volume = "06",
    pages = "025",
    year = "2015"
}

@article{King:2014xwa,
    author = {King, S. F. and M{\"u}hlleitner, M. and Nevzorov, R. and Walz, K.},
    title = "{Discovery Prospects for NMSSM Higgs Bosons at the High-Energy Large Hadron Collider}",
    eprint = "1408.1120",
    archivePrefix = "arXiv",
    primaryClass = "hep-ph",
    doi = "10.1103/PhysRevD.90.095014",
    journal = "Phys. Rev. D",
    volume = "90",
    number = "9",
    pages = "095014",
    year = "2014"
}

@article{Huang:2014cla,
    author = "Huang, Jinrui and Liu, Tao and Wang, Lian-Tao and Yu, Felix",
    title = "{Supersymmetric Subelectroweak Scale Dark Matter, the Galactic Center Gamma-Ray Excess, and Exotic Decays of the 125 GeV Higgs Boson}",
    eprint = "1407.0038",
    archivePrefix = "arXiv",
    primaryClass = "hep-ph",
    reportNumber = "FERMILAB-PUB-14-183-T",
    doi = "10.1103/PhysRevD.90.115006",
    journal = "Phys. Rev. D",
    volume = "90",
    number = "11",
    pages = "115006",
    year = "2014"
}

@inproceedings{Liu:2013gea,
    author = "Liu, Tao and Potter, C. T.",
    title = "{Exotic Higgs Decay $h \to a_1a_1$ at the International Linear Collider: a Snowmass White Paper}",
    booktitle = "{Snowmass 2013}: {Snowmass on the Mississippi}",
    eprint = "1309.0021",
    archivePrefix = "arXiv",
    primaryClass = "hep-ph",
    month = "8",
    year = "2013"
}

@article{Choi:2019yrv,
    author = "Choi, Kiwoon and Im, Sang Hui and Jeong, Kwang Sik and Park, Chan Beom",
    title = "{Light Higgs bosons in the general NMSSM}",
    eprint = "1906.03389",
    archivePrefix = "arXiv",
    primaryClass = "hep-ph",
    reportNumber = "CTPU-PTC-19-17, PNUTP-19-A11",
    doi = "10.1140/epjc/s10052-019-7473-1",
    journal = "Eur. Phys. J. C",
    volume = "79",
    number = "11",
    pages = "956",
    year = "2019"
}

@article{Yue:2025dqe,
    author = "Yue, Yuanfang and Cao, Junjie and Li, Fei and Li, Zehan",
    title = "{Attractive features of Higgsino dark matter in the next-to-minimal supersymmetric standard model}",
    eprint = "2503.10985",
    archivePrefix = "arXiv",
    primaryClass = "hep-ph",
    doi = "10.1103/hlf3-l7cp",
    journal = "Phys. Rev. D",
    volume = "112",
    number = "9",
    pages = "095022",
    year = "2025"
}

@article{Profumo:2007wc,
doi = {10.1088/1126-6708/2007/08/010},
url = {https://doi.org/10.1088/1126-6708/2007/08/010},
year = {2007},
month = {aug},
publisher = {},
volume = {2007},
number = {08},
pages = {010},
author = {Stefano Profumo and Michael J. Ramsey-Musolf and Gabe Shaughnessy},
title = {Singlet Higgs phenomenology and the electroweak phase transition},
journal = {Journal of High Energy Physics}
}

@article{Kozaczuk:2019pet,
  title = {Exotic Higgs boson decays and the electroweak phase transition},
  author = {Kozaczuk, Jonathan and Ramsey-Musolf, Michael J. and Shelton, Jessie},
  journal = {Phys. Rev. D},
  volume = {101},
  issue = {11},
  pages = {115035},
  numpages = {15},
  year = {2020},
  month = {Jun},
  publisher = {American Physical Society},
  doi = {10.1103/PhysRevD.101.115035},
  url = {https://link.aps.org/doi/10.1103/PhysRevD.101.115035}
}

@article{Carena:2019une,
  title={Electroweak phase transition with spontaneous Z 2 -breaking},
  author={ Carena, Marcela  and  Wang, Yikun  and  Carena, Marcela  and  Wang, Yikun  and  Carena, Marcela  and  Liu, Zhen },
  journal={JOURNAL of HIGH ENERGY PHYSICS},
  volume={2020},
  number={8},
  year={2020},
  month = {Aug},
  doi = {10.1007/JHEP08(2020)107},
  url = {http://dx.doi.org/10.1007/JHEP08(2020)107}
}

@article{Carena:2022yvx,
    author = "Carena, Marcela and Kozaczuk, Jonathan and Liu, Zhen and Ou, Tong and Ramsey-Musolf, Michael J. and Shelton, Jessie and Wang, Yikun and Xie, Ke-Pan",
    title = "{Probing the Electroweak Phase Transition with Exotic Higgs Decays}",
    eprint = "2203.08206",
    archivePrefix = "arXiv",
    primaryClass = "hep-ph",
    reportNumber = "FERMILAB-CONF-22-178-T",
    doi = "10.31526/lhep.2023.432",
    journal = "LHEP",
    volume = "2023",
    pages = "432",
    year = "2023"
}
\end{document}